\documentclass{general}  

\usepackage{xcolor}
\usepackage{xcolor}
\usepackage[
  colorlinks = true,
  linkcolor  = red,    
  citecolor  = blue,   
  urlcolor   = magenta
]{hyperref}
\usepackage{graphicx}
\usepackage{array}
\usepackage{tabularx}
\usepackage{txfonts}
\usepackage{lipsum}

\usepackage{titlesec}
\titlespacing{\section}{0pt}{12pt plus 2pt minus 2pt}{10pt plus 2pt minus 2pt}
\titlespacing{\subsection}{0pt}{12pt plus 2pt minus 2pt}{10pt plus 2pt minus 2pt}
\titlespacing{\subsubsection}{0pt}{12pt plus 2pt minus 2pt}{10pt plus 2pt minus 2pt}

\usepackage{float}
\usepackage{tabularx}
\usepackage{subcaption}         % necessary for continued figures, example in section 3
\usepackage{lscape}             % to rotate a single page table, example in appendix.
\usepackage{placeins}           % useful with \FloatBarrier, to keep 
\begin{document}

%%%%%%%%%%%%%%%%%%%%%%%%%%%%%%%%%%%%%%%%
% if you use custom commands in your title,
% ensure to check your title when submitting!
%%%%%%%%%%%%%%%%%%%%%%%%%%%%%%%%%%%%%%%%
   \title{Optimal subbanding to model frequency-dependent pulse profile evolution in narrowband pulsar timing}
   
   %\subtitle{Subtitle}

%%%%%%%%%%%%%%%%%%%%%%%%%%%%%%%%%%%%%%%%
% Please separate each author with the \and command
%%%%%%%%%%%%%%%%%%%%%%%%%%%%%%%%%%%%%%%%

   \author{Pratik Tarafdar\inst{1}\fnmsep\thanks{Corresponding author: pratik.tarafdar@inaf.it}
        \and Debabrata Deb\inst{2,3}
        \and Shaswata Chowdhury\inst{4}
        \and Amarnath\inst{5}
        \and Himanshu Grover\inst{6,7}
        \and Bhal Chandra Joshi\inst{7,6}
        \and Shantanu Desai\inst{8}
        \and Churchil Dwivedi\inst{9}
        \and Manjari Bagchi\inst{10,11}
        \and Jaikhomba Singha\inst{12}
        \and M. A. Krishnakumar\inst{13}
        \and K Nobleson\inst{14}
        \and Avinash Kumar Paladi\inst{15}
        \and Mayuresh Surnis\inst{16}
        \and Abhimanyu Susobhanan\inst{17}
        \and Keitaro Takahashi\inst{14}
        }

   \institute{INAF - Osservatorio Astronomico di Cagliari, via della Scienza 5, 09047 Selargius (CA), Italy
   \and Centre for Space Research, North-West University, Private Bag X6001, Potchefstroom 2520, South Africa
   \and National Institute for Theoretical and Computational Sciences (NITheCS), Potchefstroom, South Africa
   \and Radio Astronomy Centre, Tata Institute of Fundamental Research, National Centre for Radio Astrophysics, Ooty, Tamil Nadu 643001
   \and Raman Research Institute, Bengaluru 560080, Karnataka, India
   \and Department of Physics, Indian Institute of Technology Roorkee, Roorkee, Uttarakhand, 247667, India 
   \and National Centre for Radio Astrophysics, Tata Institute of Fundamental Research, SP Pune University Campus, Pune, Maharashtra, 411007, India
   \and Department of Physics, IIT Hyderabad, Kandi, Telangana 502284, India
   \and Astronomy and Astrophysics Division, Physical Research Laboratory, Thaltej Campus, Thaltej, Ahmedabad 380059, Gujarat, India
   \and The Institute of Mathematical Sciences, C. I. T. Campus, Taramani, Chennai 600113, India
   \and Homi Bhabha National Institute, Training School Complex, Anushakti Nagar, Mumbai 400094, India
   \and High Energy Physics, Cosmology and Astrophysics Theory Group (HEPCAT), Department of Mathematics and Applied Mathematics, University of Cape Town, Cape Town 7700, South Africa
   \and Radio Astronomy Centre, National Centre for Radio Astrophysics, Tata Institute of Fundamental Research, Udhagamandalam, 643001, India
   \and Faculty of Advanced Science and Technology, Kumamoto University, 2-39-1 Kurokami, Kumamoto 860-8555, Japan
   \and  Joint Astronomy Programme, Department of Physics, Indian Institute of Science, Bengaluru, Karnataka, 560012, India
   \and Department of Physics, IISER Bhopal, Bhauri Bypass Road, Bhopal, 462066, Madhya Pradesh, India
   \and School of Physics, Indian Institute of Science Education and Research Thiruvananthapuram, Maruthamala PO, Thiruvananthapuram, Kerala 695551, India}

   \date{}

% \abstract{}{}{}{}{}
% 5 {} token are mandatory
 
  \abstract
    {}
    {Frequency-dependent (FD) parameters are widely used in high-precision narrowband pulsar timing to account for apparent delays due to intrinsic frequency-dependent pulse profile evolution across wide radio-frequency bands. While such parameters, corresponding to the coefficients of a polynomial in logarithm of the radio-frequency, lead to whitening of the timing residuals, these can inherently absorb delays such as chromatic delays due to the ionised interstellar medium (IISM). Our aim is to study schemes to potentially eliminate the requirement of FD parameters with appropriate template-making and subband optimisation techniques in pulsar timing.} 
    {We investigate two separate schemes of template generation -- frequency-averaged and frequency-resolved -- for the estimation of pulse times-of-arrival (ToAs), and perform quantitative tests on a simulated pulse profile evolving across the 300$-$500 MHz radio-frequency band. These tests utilise Bayesian Information Criterion (BIC) and F-scores to probe the statistical significance of FD parameter fits in a timing model as function of the number of subbands using both varieties of templates.} 
    {Our results demonstrate, as a proof of principle, that frequency-resolved templates inherently mitigate timing delays arising from profile-frequency evolution. Frequency-averaged templates can achieve a similar degree of compensation through judicious subbanding, provided that the number of subbands is optimally chosen according to the nature and magnitude of profile evolution in individual pulsars.} 
    {}


   \maketitle

%%%%%%%%%%%%%%%%%%%%%%%%%%%%%%%%%%%%%%%%%%%%%%%%%%%%%%%%%%%%%%
\section{Introduction}
\label{sec:intro}
\noindent
In high-precision pulsar timing, the frequency-dependent (FD) parameters are a phenomenological way of modeling systematic apparent timing delays caused primarily by pulse profile evolution across the observing radio band. The fundamental idea is that different radio frequencies are emitted from different regions of the pulsar magnetosphere, resulting in the shape of the pulse profile changing as a function of observing frequency \citep{Cordes78}. This makes it difficult to map a point in the pulse profile observed at a certain radio frequency to a longitude on the pulsar’s surface. Upon using a single frequency-averaged template to determine the pulse times-of-arrival (ToAs), the measured arrival times exhibit a smooth systematic FD delay or offset that is unrelated to interstellar dispersion or scattering. Such delays are usually expressed using a polynomial expression in the logarithm of the radio frequency \citep{ZSD15,ABB16,SHML21} --
\begin{equation}
\label{eqn:FD}
\Delta t_{FD}=\sum^{n}_{i=1}c_{i}\left[\text{log}\left(\frac{\nu}{\text{1 GHz}}\right)\right]^{i}
\end{equation} 
\noindent
where, $\Delta t_{FD}$ is the delay due to profile-frequency evolution, $\nu$ is the observing frequency and $c_i$ are the coefficients (in dimensions of time) referred to as the FD parameters. Pulse profile evolution, though typically smooth over wide bandwidths, rarely follows a simple power law. Hence, the logarithmic definition provides a flexible empirical description without requiring a physical emission model, capturing effects such as changing relative amplitudes of profile components, varying component widths, shifts in the effective pulse centroid, or template mismatch across telescopes. We also note in passing that different pulsar timing array (PTA) experiments may use templates generated through different data processing pipelines. This sometimes requires introducing system-dependent FD parameters, known as FDJUMPs, while combining data from different pulsar timing campaigns \citep{Susobhanan24}.  

However, it is important to note that such FD terms can become covariant with frequency-dependent effects such as dispersion, scattering and other chromatic variations \citep{Alam21}. FD parameters are not intended to absorb propagation effects in the interstellar medium (ISM), but to account for intrinsic or template-related profile evolution. Wideband templates (named \textit{portraits}, reconstructed from the significant principal components) \citep{pdr14, Pennucci19} model profile-frequency evolution by construction, thus rendering inclusion of FD parameters unnecessary \citep{Nobleson22, Paladi24, Susobhanan26}. However, PTAs have reported FD parameter fits in their wideband timing solutions \citep{Agazie23, Curylo23}. FD parameters have also been included in timing subbanded ToAs generated with wideband portraits \citep{Miles23, Miles25}. While fitting FD parameters in such cases could lead to whitening of the residuals, it imminently runs into the risk of absorbing a significant fraction of chromatic delays that do not originate from intrinsic profile-frequency evolution, and require careful modeling. The present age of ultra-wideband receivers and sub-$\mu$s to nanosecond timing precisions necessitates alternatives to the present FD formalism. Results have been reported in the direction of mitigating the necessity of FD parameters with the use of frequency-resolved templates coupled with \textit{pulsar-specific subbanding} for ToA estimation \citep{Tarafdar22, Rana25}. Our work presents a rigorous statistical validation of such relations between template generation or subbanding techniques and the significance of FD parameter fits. The prevalent FD parameterisation is fundamentally phenomenological, modelling chromatic delays as polynomials in the logarithm of observing frequencies. Several studies have argued that this approach should be superseded by physically motivated models that explicitly describe pulse-profile evolution and frequency-dependent propagation effects through the ISM \citep{css16, Lentati17,sc17, sofia26}. 

In this work, we use simulations to initiate a series of investigations for comparing the significances of fitting FD parameters to ToAs generated using frequency-averaged and frequency-resolved templates. We also put to test, the hypothesis of optimal subbanding, where steepness of evolution determines the number of subbands beyond which FD parameter fits become statistically insignificant. Section \ref{sec:simulations} describes the simulation setup and the dataset. Section \ref{sec:template_toas} elaborates the methods of template generation and ToA estimation. Section \ref{sec:timing} explains the scheme of fitting FD parameters that has been adopted in this work. Section \ref{sec:model_selection} reports the statistical significance of FD fits to timing residuals from various combinations of template and subband configurations. Section \ref{sec:profile_domain} validates such statistical optimisation using direct visualisation of subbanded pulses in the profile domain. In section \ref{sec:conclusion}, we summarise our conclusions and outline the future direction of study.
%%%%%%%%%%%%%%%%%%%%%%%%%%%%%%%%%%%%%%%%%%%%%%%%%%%%%%%%%%%%%%
\section{Simulations \& dataset}
\label{sec:simulations}
\noindent
For this study, we generated synthetic pulsar observations under controlled conditions, using a dedicated simulation framework developed in-house as part of the Indian Pulsar Timing Array (InPTA) \citep{Joshi18, Joshi22, Tarafdar22, Rana25} efforts. The simulator is designed to generate \texttt{PSRFITS} archives \citep{HVM04} while allowing the physical and instrumental effects to be tuned independently. Since development of the simulation framework constitutes substantial effort that we plan to report separately, in the context of the present work, we restrict ourselves to a description of the simulation setup and characteristics of the used dataset.

The simulations use a synthetic source designation J0000+0000 to generate archives as a controlled test case in which the profile-evolution properties are prescribed explicitly. This designation is introduced as a neutral identifier for the simulated source and does not correspond to the observed properties of any real pulsar. The simulation is configured to specifically represent the observing characteristics of the InPTA experiment with Band 3 of the upgraded Giant Metrewave Radio Telescope (uGMRT) \citep{Gupta17}, having a frequency range of $300-500$ MHz. Each simulated observation has a sampling time of 5.12 $\mu$s, for an effective phase-resolution of 2048 bins, with 150 sub-integrations, each of 10s durations, and 128 frequency sub-bands. The channels are ordered in a lower-sideband (LSB) configuration \citep{Reddy17}. 

For our present study, intrinsic pulse emission is represented by a single Gaussian component. A frequency-dependent asymmetric tail is introduced in order to incorporate frequency evolution of the pulse width. The tail is parameterised by a characteristic timescale that follows a power-law dependence on observing frequency, with a reference value of 8 phase bins at 1\,GHz and a frequency scaling index of $-4$. For the adopted 5.12\,$\mu$s sampling interval, the width changes from $\sim$ 0.66\,ms at 500\,MHz to 1.60\,ms at 400\,MHz and 5.06\,ms at 300\,MHz. This frequency-dependent asymmetric broadening thus introduces a width-evolution across the observing band while retaining a single intrinsic emission component. It is important to note that such an asymmetric tail is also the characteristic feature of scatter-broadened pulsars, e.g. PSR J1643$-$1224. PTA data releases report 5 \citep{Agazie23}, 3 \citep{Zic23}, and 2 \citep{Miles25} FD parameter fits for PSR J1643$-$1224. Within the simulational framework given for our present work, the tail should be treated as frequency evolution of the intrinsic profile.  

The dispersion measure (DM) is held fixed throughout the simulation. We refrain from introducing stochastic DM variations, solar-wind contribution or DM jumps in order to provide a controlled environment for exclusively studying the effects of intrinsic profile evolution on template construction and timing solutions. The simulated profiles include additive injected radiometer noise that is not explicitly modulated by a frequency-dependent system-temperature model. Flux per channel varies across the band following a spectral-index prescription. The resulting configuration isolates effects of the simulated profile evolution from any additional frequency- or time-dependent perturbation. 

\begin{table}[htpb!]
    \centering
    \caption{Parameters used in simulation}
    \label{tab:sim_config}\hspace*{-0.25cm}
    \begin{tabularx}{0.45\textwidth} { 
  >{\arraybackslash}X 
  >{\arraybackslash}X }
    \hline\hline
    {\footnotesize Parameter} & {\footnotesize Value} \\
    \hline
    {\footnotesize Source designation} & {\footnotesize J0000+0000} \\
    {\footnotesize Central frequency} & {\footnotesize 400 MHz} \\
    {\footnotesize Bandwidth} & {\footnotesize 200 MHz} \\
    {\footnotesize Number of channels} & {\footnotesize 128} \\
    {\footnotesize Phase bins} & {\footnotesize 2048} \\
    {\footnotesize Subintegrations} & {\footnotesize 150} \\
    {\footnotesize Total integration time} & {\footnotesize 1500 s} \\
    {\footnotesize Sampling time} & {\footnotesize 5.12 $\mu$s} \\
    {\footnotesize Pulse period} & {\footnotesize 16.052 ms} \\
    {\footnotesize DM} & {\footnotesize 9.0018364 pc cm$^{-3}$} \\
    {\footnotesize No. of profile components} & {\footnotesize 1} \\
    {\footnotesize MJD start} & {\footnotesize 59500} \\
    {\footnotesize MJD end} & {\footnotesize 60600} \\
    {\footnotesize Cadence} & {\footnotesize 14 days} \\
    {\footnotesize Tail reference scale} & {\footnotesize 8 bins at 1 GHz} \\
    {\footnotesize Tail frequency index} & {\footnotesize $-4$} \\
    {\footnotesize Noise model} & {\footnotesize Frequency-independent} \\
    {\footnotesize Sideband configuration} & {\footnotesize LSB} \\
    {\footnotesize Flux treatment} & {\footnotesize FD power-law scaling} \\
    {\footnotesize Spectral-index} & {\footnotesize $-1.6$} \\
    \hline\hline
    \end{tabularx}
\end{table}
\begin{figure}[htpb!]
    \centering \includegraphics[width=0.5\textwidth,clip,trim=1.5cm 3cm 1.5cm 5.5cm]{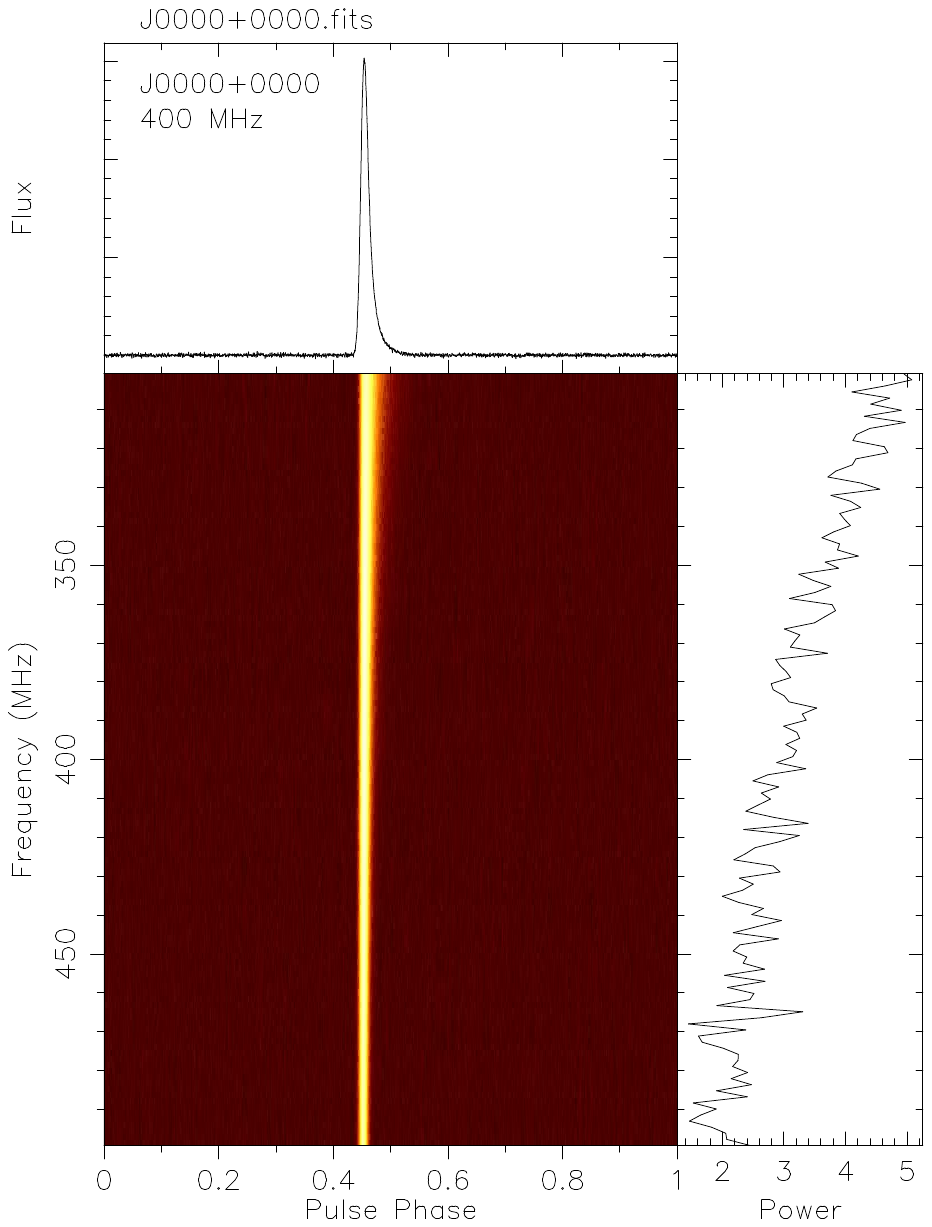}
    \caption{Profile of the simulated pulsar J0000+0000 across uGMRT Band 3. Central panel depicts frequency-resolved pulse intensity as a function of frequency and phase. Upper panel shows frequency-integrated flux across phase bins. Panel on the right displays phase-integrated flux distribution across channels.
    }
    \label{fig:simulated_profile}
\end{figure}

The principal parameters of simulation are summarised in Table~\ref{tab:sim_config}, and a representative epoch of the simulated pulse is depicted in Figure~\ref{fig:simulated_profile}. The central panel in the figure shows frequency-resolved pulse intensity. The upper and right panels show the corresponding frequency-integrated profile and power distributions, respectively. While the asymmetric tail produces an effective increase in the observed pulse width towards lower frequencies, the global spectral index introduces a frequency-dependent flux-density scaling.

%%%%%%%%%%%%%%%%%%%%%%%%%%%%%%%%%%%%%%%%%%%%%%%%%%%%%%%%%%%%%%

\section{Template construction and ToA generation}
\label{sec:template_toas}
\subsection{Frequency-resolved and frequency-collapsed templates}
\label{ssec:template}
\begin{figure}[h!]
\centering
\includegraphics[width=0.5\textwidth,clip,trim=1.5cm 3cm 1.5cm 5.5cm]{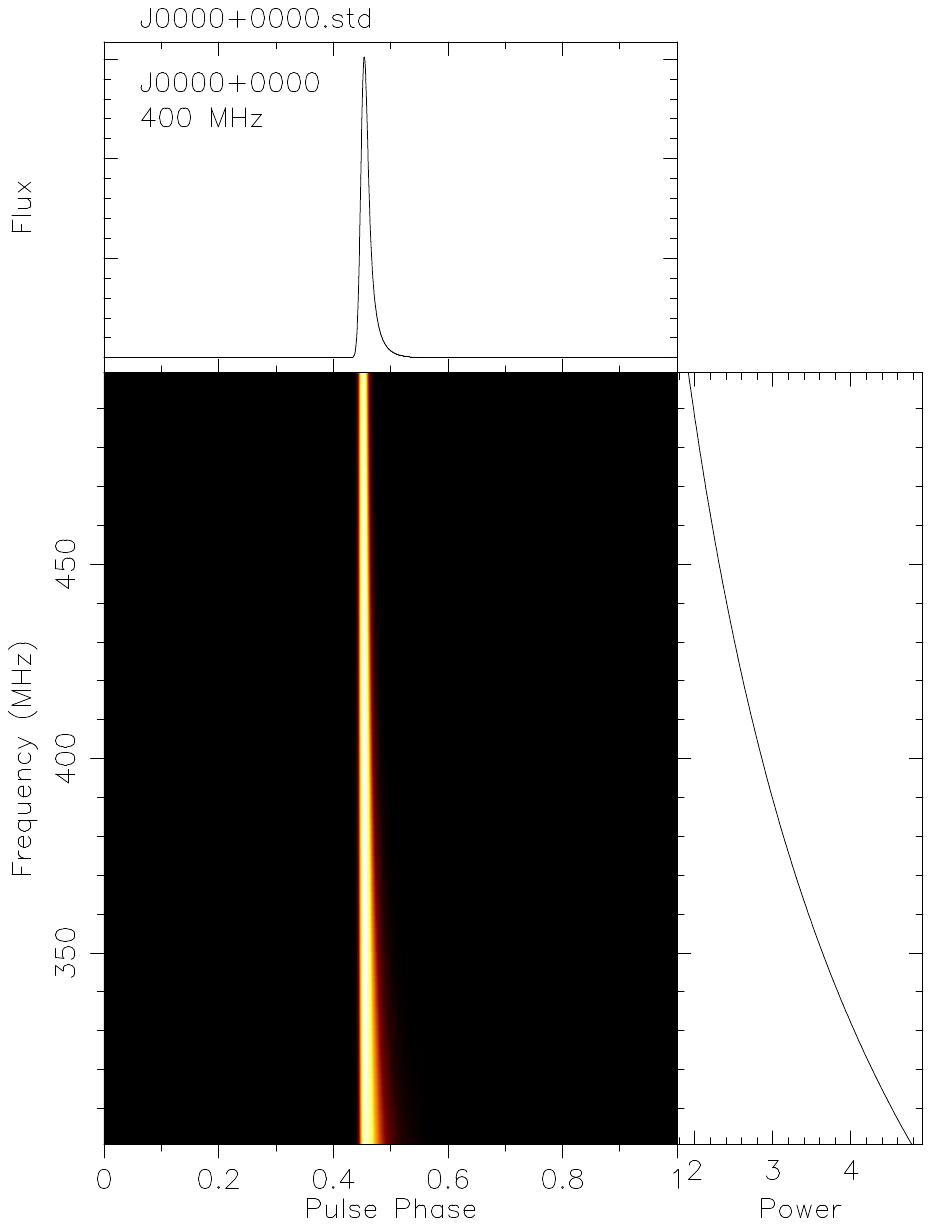}
\caption{Processed noise-free model profile.}
\label{fig:template}
\end{figure}

\noindent
A single-epoch model profile is simulated using the set-up described in section \ref{sec:simulations}. The sole exception is the injection of zero radiometer noise for the model-epoch in order to obtain the cleanest pulse with the highest possible signal-to-noise ratio (S/N). The model profile is then \textit{processed}, i.e. a) dispersion-corrected with the known injected constant DM listed in table~\ref{tab:sim_config}, b) time-averaged, and c) frequency-reversed since the simulator generates negative bandwidth in concurrence with uGMRT sideband configuration used for the InPTA experiment. Figure~\ref{fig:template} depicts a processed noise-free model profile used as a template for cross-correlation with the archives to generate the ToAs. 
This processed profile is finally used to construct two sets of templates. The first set consists of a single template with full frequency and phase resolutions (FR henceforth). The second set of templates preserves full phase resolution with the entire bandwidth progressively divided into $1$, $2$, $4$, $8$, $16$, $32$, $64$ and $128$ frequency-averaged components respectively, where $1$ component corresponds to a completely \textit{frequency-collapsed} (using \texttt{pam} module of \texttt{PSRCHIVE} \citep{HVM04}) or FC template (FC$_{1}$ henceforth), while $2$, $4$, $8$, $16$, $32$ and $64$ components correspond to $2$ $\times$ $100$ MHz FC templates (FC$_{2}$), $4$ $\times$ $50$ MHz FC templates (FC$_{4}$), $8$ $\times$ $25$ MHz FC templates (FC$_{8}$), $16$ $\times$ $12.5$ MHz FC templates (FC$_{16}$), $32$ $\times$ $6.25$ MHz FC templates (FC$_{32}$), and $64$ $\times$ $3.125$ MHz FC templates (FC$_{64}$). $128$ components (FC$_{128}$) correspond to a fully frequency-resolved template (identical to FR). The objective behind such construction will be clear when we fit FD parameters to the ToA residuals obtained upon cross-correlation of archives with a set of FC$_n$ templates.

\subsection{ToA estimation}
\label{ssec:toas}
\noindent
All the ToAs described below are generated by cross-correlating simulated templates and archives using the \textit{Fourier domain with Markov chain Monte Carlo} (\texttt{FDM}) algorithm incorporated in the \texttt{pat} module of \texttt{PSRCHIVE} \citep{HVM04}.
\subsubsection{Subbanded ToAs from FR template}
\label{sssec:toas_fr}
\noindent
The first category of ToAs, generated with the FR template, consists of $4$, $8$, $16$, $32$, $64$ and $128$ subbanded sets of ToAs for each of the $79$ epochs in the simulated dataset. 
\subsubsection{Subbanded ToAs from FC$_n$ templates}
\label{sssec:toas_fcn}
\noindent
The second category consists of $8$ sets, each comprising $128$ ToAs per epoch. Set $1$ contains $128$ ToAs estimated by cross-correlating $128$-channel archives with the FC$_{1}$ template. Set $2$ of ToAs is generated by cross-correlating-- 
\begin{enumerate}
    \item the lower $64$ channels of archives (within $300-400$ MHz) with the first component of FC$_{2}$ averaged across $300-400$ MHz.
    \item the upper $64$ channels of archives (within $400-500$ MHz) with the second component of FC$_{2}$ averaged across $400-500$ MHz.
\end{enumerate}
Set 3 is generated by cross-correlating--
\begin{enumerate}
    \item $32$ channels of the archives (within $300-350$ MHz) with the first component of FC$_{4}$ averaged across $300-350$ MHz.
    \item $32$ channels of the archives (within $350-400$ MHz) with the second component of FC$_{4}$ averaged across $350-400$ MHz.
    \item $32$ channels of the archives (within $400-450$ MHz) with the third component of FC$_{4}$ averaged across $400-450$ MHz.
    \item $32$ channels of the archives (within $450-500$ MHz) with the fourth component of FC$_{4}$ averaged across $450-500$ MHz, and so on.
\end{enumerate}
In accordance with the workflow enumerated above, set $m$ of ToAs ($m \in \{1,8\}$) is constructed by cross-correlating archives partially-collapsed to $128/n$ channels with the corresponding component of FC$_n$ averaged across its respective frequency band, where $n \in \{1,2,4,8,16,32,64,128\}$. 

In the following sections, we present the results of FD parameter fits to timing residuals corresponding to the different sets of ToAs described above. We then proceed to compare statistical significances of the fits, thus investigating the utility of FD parameters in each case. 

%%%%%%%%%%%%%%%%%%%%%%%%%%%%%%%%%%%%%%%%%%%%%%%%%%%%%%%%%%%%%%
\section{Timing with FD parameters}
\label{sec:timing}
\noindent
We fit an appropriate delay model to the residuals obtained from the generated ToAs using the \texttt{TEMPO2} pulsar timing package \citep{HEM06, Edwards06}. Since the archives are simulated from a well-constrained pulsar ephemeris, we do not fit for the spin, astrometric, binary and ISM-related parameters. We rather try to determine an optimal combination of FD parameters to model delays due to the simulated evolution of pulse profile with the radio-frequency. The FD parameters, as defined in equation \ref{eqn:FD}, correspond to coefficients of a polynomial expression in the logarithm of frequency $\nu$ to represent delays in the times of arrival of pulses owing to the variation of the pulsar's intrinsic emission across the radio bandwidth. In the present analysis, we fit different combinations of successive $c_i$ (FD$_i$, henceforth) till $i=5$, considering that the cut-off at the 5th order is a judicious choice based on the required numbers of FD parameters reported by global PTA experiments. Combinations are chosen among the coefficients of leading -- a) odd powers, b) even powers, and c) all consecutive powers of equation \ref{eqn:FD}. Table~\ref{table:0} lists all such possible FD combinations. In the following section, we present a scheme for selection of the best FD models from the table for $4$, $8$, $16$, $32$, $64$, $128$ subbanded ToAs per epoch generated from FR and FC$_n$ templates. We shall refer to FD$_{i}$ as $i$; combination of FD$_{i}$ and FD$_{j}$ as $\left[i,j\right]$; combination of consecutive FDs from $i$ to $j$ as $\left[i-j\right]$ etc.; for the sake of brevity. 

\begin{table}[ht!]
\caption{Choice of FD combinations}             
\label{table:0}    
\centering     
\begin{tabularx}{0.45\textwidth} { 
  >{\centering\arraybackslash}X 
  >{\centering\arraybackslash}X 
  >{\centering\arraybackslash}X }      
\hline\hline               
{\footnotesize Series} & {\footnotesize \#Leading powers} & {\footnotesize Combination} \\         
\hline                      
   {\footnotesize Full} & {\footnotesize 1} & {\scriptsize FD$_{1}$} \\
    & {\footnotesize 2} & {\scriptsize FD$_{1}$+FD$_{2}$} \\
    & {\footnotesize 3} & {\scriptsize FD$_{1}$+FD$_{2}$+FD$_{3}$} \\
    & {\footnotesize 4} & {\scriptsize FD$_{1}$+FD$_{2}$+FD$_{3}$+FD$_{4}$} \\
    & {\footnotesize 5} & {\scriptsize FD$_{1}$+FD$_{2}$+FD$_{3}$+FD$_{4}$+FD$_{5}$} \\
\hline
    {\footnotesize Odd} & {\footnotesize 1} & {\scriptsize FD$_{1}$} \\    
    & {\footnotesize 2} & {\scriptsize FD$_{1}$+FD$_{3}$} \\
    & {\footnotesize 3} & {\scriptsize FD$_{1}$+FD$_{3}$+FD$_{5}$} \\
\hline
   {\footnotesize Even} & {\footnotesize 1} & {\scriptsize FD$_{2}$} \\
    & {\footnotesize 2} & {\scriptsize FD$_{2}$+FD$_{4}$} \\
\hline
\end{tabularx}
\end{table} 

%%%%%%%%%%%%%%%%%%%%%%%%%%%%%%%%%%%%%%%%%%%%%%%%%%%%%%%%%%%%%%
\section{FD model selection \& subband optimisation from fit statistics}
\label{sec:model_selection}

\subsection{Subbanded ToAs from FR template}
\label{ssec:model_selection_fr}

\subsubsection{FD model selection for a given subband}
\label{sssec:model_selection_fr_bic}
\noindent
In the first category of ToAs (as explained in subsection \ref{sssec:toas_fr}), sets with $4$, $8$, $16$, $32$, $64$ and $128$ ToAs per epoch have different sample sizes. We use the Bayesian Information Criterion (BIC)~\citep{Schwarz78}, also known as Schwarz Information Criterion, for model selection among the finite set listed in table~\ref{table:0} for each subbanded set of ToAs. The figure of merit imposes penalties proportional to the number of parameters in order to counter an increase in maximum likelihood caused by possible overfitting. The Akaike Information Criterion (AIC)~\citep{Akaike78} imposes similar penalties but BIC is a more robust measure for large and varying sample sizes, which is appropriate in the present context~\citep{Liddle07}. The BIC is formally defined as,
\begin{eqnarray}
    BIC&=&k \ln(N)-2 \ln(\hat{L}) \nonumber \\
    &=&k \ln(N)+\chi^2
\label{eqn:bic}
\end{eqnarray}
where, $k$ is the number of fit parameters, $N$ is the sample size (number of ToA residuals), and $\hat{L}$ is the maximum value of the likelihood function for a given model. Assuming an independent Gaussian distribution of the model errors and an extremised derivative of the log likelihood with respect to the true variance, this maximised likelihood function assumes a value in terms of the deviance ($\chi^2$) statistic, where $\ln(\hat{L})$ becomes $-\chi^2/2$. 
The best model corresponds to the minimum value of BIC for a given sample. Based on the above definitions, we select the optimal FD models for various subband configurations obtained from a frequency-resolved (FR) template. A similar comparison between different subbanded sets of ToAs cannot be achieved with BIC because of non-uniform sample size (8-subbanded ToAs are double the number of 4-subbanded ToAs, and so on). An appropriate F-statistic and its corresponding p-value at a given significance are used for such comparision (explained in the following subsection). The BIC values for all FD model--subband pairs are provided in table~\ref{tab:fd_model_select_fr}. Entries marked as `$-$' signify models for which the fit diverges, i.e. post-fit root mean square (RMS) residual is greater than the pre-fit RMS residual. Reported numbers clearly indicate that \textit{`No FD'} model (when no FD parameters are fitted) attains the minimum values of BIC (marked in boldface) for all subband configurations obtained from FR template. We also note that the difference in BICs ($\Delta$BIC) between the minimum and its nearest value for a given subband is always less than 10. This implies that FD fits are statistically inconsequential when compared to no FD fits in all cases.

\begin{table*}[ht!]
\caption{Model selection using BIC (FR template). Boldfaces mark the minimum BICs for a given subband. `-' marks divergent fits. $\Delta$ToAs are medians of ToA uncertainties across a given subband.}                 
\label{tab:fd_model_select_fr}    
\centering     
\begin{tabularx}{\textwidth} { 
  >{\centering\arraybackslash}X
  >{\centering\arraybackslash}X
  >{\centering\arraybackslash}X
  >{\centering\arraybackslash}X
  >{\centering\arraybackslash}X
  >{\centering\arraybackslash}X
  >{\centering\arraybackslash}X }      
\hline\hline               
{\footnotesize FD Model} & {\footnotesize 4 subbands} & {\footnotesize 8 subbands} & {\footnotesize 16 subbands} & {\footnotesize 32 subbands} & {\footnotesize 64 subbands} & {\footnotesize 128 subbands} \\
 & {\footnotesize $\Delta$ToA= 0.2 $\mu$s} & {\footnotesize $\Delta$ToA= 0.3 $\mu$s} & {\footnotesize $\Delta$ToA= 0.4 $\mu$s} & {\footnotesize $\Delta$ToA= 0.6 $\mu$s} & {\footnotesize $\Delta$ToA= 0.8 $\mu$s} & {\footnotesize $\Delta$ToA= 1.2 $\mu$s} \\
\hline                      
   {\footnotesize 1} & {\footnotesize 330.7615} & {\footnotesize 613.4978} & {\footnotesize 1275.9041} & {\footnotesize 2608.4904} & {\footnotesize 5277.7867} & {\footnotesize 10674.0630} \\    
   {\footnotesize $\left[1-2\right]$} & {\footnotesize 333.0472} & {\footnotesize 613.9667} & {\footnotesize 1278.4061} & {\footnotesize 2610.2756} & {\footnotesize 5281.4450} & {\footnotesize 10678.2244} \\
   {\footnotesize $\left[1-3\right]$} & {\footnotesize 336.4930} & {\footnotesize 618.9956} & {\footnotesize 1283.2181} & {\footnotesize 2616.5107} & {\footnotesize 5287.8433} & {\footnotesize 10685.9459} \\
   {\footnotesize $\left[1-4\right]$} & {\footnotesize --} & {\footnotesize 624.9644} & {\footnotesize 1289.6302} & {\footnotesize 2623.7959} & {\footnotesize 5296.1817} & {\footnotesize 10694.9574} \\
   {\footnotesize $\left[1-5\right]$} & {\footnotesize --} & {\footnotesize 630.9433} & {\footnotesize 1296.5322} & {\footnotesize 2631.2611} & {\footnotesize 5304.4800} & {\footnotesize 10703.7489} \\
   {\footnotesize $\left[1,3\right]$} & {\footnotesize 333.2272} & {\footnotesize 614.2267} & {\footnotesize 1278.7061} & {\footnotesize 2610.5656} & {\footnotesize 5281.7550} & {\footnotesize 10678.4844} \\
   {\footnotesize $\left[1,3,5\right]$} & {\footnotesize 336.4930} & {\footnotesize 619.1756} & {\footnotesize 1283.5081} & {\footnotesize 2616.7207} & {\footnotesize 5288.0033} & {\footnotesize 10686.0759} \\
   {\footnotesize 2} & {\footnotesize 330.4115} & {\footnotesize 613.0378} & {\footnotesize 1275.4941} & {\footnotesize 2608.0004} & {\footnotesize 5277.3467} & {\footnotesize 10673.6630} \\
   {\footnotesize $\left[2,4\right]$} & {\footnotesize 333.5572} & {\footnotesize 614.7367} & {\footnotesize 1279.2661} & {\footnotesize 2611.1456} & {\footnotesize 5282.3250} & {\footnotesize 10678.9944} \\
   {\footnotesize No FD} & {\footnotesize \textbf{327.1957}} & {\footnotesize \textbf{608.7489}} & {\footnotesize \textbf{1270.4720}} & {\footnotesize \textbf{2602.5152}} & {\footnotesize \textbf{5271.0783}} & {\footnotesize \textbf{10666.3115}} \\
\hline                                 
\end{tabularx}
\end{table*}

\subsubsection{Subband optimisation from F-statistics}
\label{sssec:subband_optimisation_fr_fstatistic}
\noindent
While FD model selection across samples with the same number of ToAs (equal subbands in case of cross-correlation with a FR template) can be done by minimising BIC, the determination of optimal number of subbands is achieved by defining an appropriate F-statistic, given by \citep{Bevington03}
\begin{equation}
    \label{eqn:f-value}
    F=\left.\frac{\left(\chi^2_{noFD}-\chi^2_{FD}\right)}{df_{noFD}-df_{FD}}\middle/\frac{\chi^2_{FD}}{df_{FD}}\right.=\frac{\left(\chi^2_{noFD}-\chi^2_{FD}\right)}{\chi^2_{FD}}\frac{\left(N-1-k\right)}{k}
\end{equation}
where $N$ is the number of ToAs, $k$ is the number of fitted FD parameters, $df_{noFD}$ and $df_{FD}$ are the numbers of degrees of freedom for `No FD' and FD fit models respectively, $\chi^2_{noFD}$ and $\chi^2_{FD}$ correspond to fitted $\chi^2$s for `No FD' and FD-fit models respectively. The \textit{`F-ratio'} is defined as $F/F_c$, where $F_c$ is the critical F-value obtained from F-distribution tables at a desired statistical significance $\alpha$ (chosen to be $0.05$, irrespective of the choice of subbands being used) and for a given combination of $df_{noFD}$ and $df_{FD}$. Then, p-values of the calculated F-ratios are computed using survival functions (SF), i.e. complement of the corresponding cumulative distribution functions (CDF), implemented in the statistical library of \texttt{SciPy} \citep{Virtanen20}. It gives the upper-tail probability of the calculated F-statistic. The p-values for all FD model--subband pairs are provided in table~\ref{tab:subband_opt_fr}. Those marked `$-$', as mentioned in subsection \ref{sssec:model_selection_fr_bic}, signify FD models for which the fit diverges, i.e. post-fit RMS residual is greater than the pre-fit RMS residual. The study shows that p-values never drop below the significance level $\alpha$ for subband configurations obtained with a FR template. Thus, we fail to reject the \textit{null hypothesis} which, in our case, states that any difference between FD and `No FD' model is due to random chance. It essentially means that, for FR templates, FD fits are indistinguishable from a `No FD' model in terms of statistical significance. We shall discuss the implications of this finding in detail upon comparision with results for subbanded ToAs obtained using FC$_n$ templates in the following sections. \\

\begin{table*}[ht!]
\caption{Subband optimisation using p-values from F-statistic (FR template).  `-' marks divergent FD fits. $\Delta$ToAs are medians of ToA uncertainties across a given subband.}       
\label{tab:subband_opt_fr}    
\centering     
\begin{tabularx}{\textwidth} { 
  >{\centering\arraybackslash}X
  >{\centering\arraybackslash}X
  >{\centering\arraybackslash}X
  >{\centering\arraybackslash}X
  >{\centering\arraybackslash}X
  >{\centering\arraybackslash}X
  >{\centering\arraybackslash}X }      
\hline\hline               
{\footnotesize  FD Model} & {\footnotesize  4 subbands} & {\footnotesize  8 subbands} & {\footnotesize  16 subbands} & {\footnotesize  32 subbands} & {\footnotesize  64 subbands} & {\footnotesize  128 subbands} \\ 
 & {\footnotesize $\Delta$ToA= 0.2 $\mu$s} & {\footnotesize $\Delta$ToA= 0.3 $\mu$s} & {\footnotesize $\Delta$ToA= 0.4 $\mu$s} & {\footnotesize $\Delta$ToA= 0.6 $\mu$s} & {\footnotesize $\Delta$ToA= 0.8 $\mu$s} & {\footnotesize $\Delta$ToA= 1.2 $\mu$s} \\ 
\hline                      
   {\footnotesize  1} & {\footnotesize  0.4563} & {\footnotesize  0.7417} & {\footnotesize  0.5051} & {\footnotesize  0.4925} & {\footnotesize  0.5000} & {\footnotesize  0.5469} \\    % inserting body of the table
   {\footnotesize $\left[1-2\right]$} & {\footnotesize  0.3966} & {\footnotesize  0.2601} & {\footnotesize  0.3465} & {\footnotesize  0.2762} & {\footnotesize  0.3421} & {\footnotesize  0.3556} \\
   {\footnotesize  $\left[1-3\right]$} & {\footnotesize  0.3912} & {\footnotesize  0.2993} & {\footnotesize  0.3429} & {\footnotesize  0.3133} & {\footnotesize  0.3542} & {\footnotesize  0.4034} \\
   {\footnotesize  $\left[1-4\right]$} & {\footnotesize  --} & {\footnotesize  0.3745} & {\footnotesize  0.4103} & {\footnotesize  0.3884} & {\footnotesize  0.4556} & {\footnotesize  0.5096} \\
   {\footnotesize  $\left[1-5\right]$} & {\footnotesize  --} & {\footnotesize  0.4458} & {\footnotesize  0.4992} & {\footnotesize  0.4681} & {\footnotesize  0.5482} & {\footnotesize  0.5912} \\
   {\footnotesize  $\left[1,3\right]$} & {\footnotesize  0.4086} & {\footnotesize  0.2724} & {\footnotesize  0.3644} & {\footnotesize  0.2896} & {\footnotesize  0.3596} & {\footnotesize  0.3705} \\
   {\footnotesize  $\left[1,3,5\right]$} & {\footnotesize  0.3912} & {\footnotesize  0.3084} & {\footnotesize  0.3586} & {\footnotesize  0.3235} & {\footnotesize  0.3627} & {\footnotesize  0.4109} \\
   {\footnotesize  2} & {\footnotesize  0.4221} & {\footnotesize  0.4435} & {\footnotesize  0.4580} & {\footnotesize  0.4404} & {\footnotesize  0.4523} & {\footnotesize  0.4968} \\
   {\footnotesize  $\left[2,4\right]$} & {\footnotesize  0.4316} & {\footnotesize  0.2981} & {\footnotesize  0.4002} & {\footnotesize  0.3183} & {\footnotesize  0.3940} & {\footnotesize  0.4017} \\
\hline                                  
\end{tabularx}
\end{table*}

\subsection{Subbanded ToAs from FC$_n$ templates}
\label{sec:model_selection_fcn}

\subsubsection{FD model selection for a given subband}
\label{sssec:model_selection_fcn_bic}
\noindent
The procedure of FD model selection for sets of ToAs obtained from the FC$_n$ templates is identical to that for ToAs obtained from FR templates. The only difference is that the total number of ToAs generated from FC$_n$ templates in the $300$-$500$ MHz band remains the same across all subband configuration (128 ToAs $\times$ 79 epochs $=$ 10112 ToAs for all FC$_n$ where $n \in \{2,4,8,16,32,64,128\}$). This permits the use of BIC not only for FD model selection but also for subband optimisation in addition to F-ratios and their corresponding p-values. In this work, we compare FR and FC$_n$ results $n=4$ onwards as fitting FD trend requires more than 2 subbands across the operational bandwidth. The BIC values for all FD model-subband pairs from FC$_n$ templates are listed in table \ref{tab:fd_model_select_fcn}. Numbers in boldface indicate the minimum BICs for a given number of subbands. It is observed that the least value of BIC is attained by the $\left[1-5\right]$ model for $4-16$ subband configurations. $\Delta$BIC is observed to be greater than $10$ between different FD models for all such cases, thus substantiating the statistical significance of the preferred models. The situation starts to evolve beyond $32$ subbands when BIC attains its least value for $\left[2,4\right]$ model which, though preferred, becomes statistically indistinguishable from most of the other models ($\Delta$BIC $<10$) \textit{signifying a transition from FD-significance to FD-indifference}. The transition is complete at $64$ subbands where the minimum BIC corresponds to a `No FD' model and FD parameters become insignificant altogether. Since the size of dataset (no. of ToAs) is uniform across all subbanded configurations for FC$_n$ templates, a comparision of the minimum BICs across subbands is also permissible as stated earlier. It shows that the optimal number of subbands (just enough to make FD parameters insignificant) is achieved at $64$ subbands. In the subsequent sections, this optimisation is further established with F-ratios (corresponding p-values) and a direct inspection of residual profiles across the adjacent subbands.
\begin{table*}[ht!]
\caption{Model selection using BIC (FC$_n$ templates). Boldfaces signify the minimum BICs for a given subband. $\Delta$ToAs are medians of ToA uncertainties across a given subband.}       
\label{tab:fd_model_select_fcn}    
\centering     
\begin{tabularx}{\textwidth} { 
  >{\centering\arraybackslash}X
  >{\centering\arraybackslash}X
  >{\centering\arraybackslash}X
  >{\centering\arraybackslash}X
  >{\centering\arraybackslash}X
  >{\centering\arraybackslash}X
  >{\centering\arraybackslash}X }      
\hline\hline               
{\footnotesize  FD Model} & {\footnotesize  4 subbands} & {\footnotesize  8 subbands} & {\footnotesize  16 subbands} & {\footnotesize  32 subbands} & {\footnotesize  64 subbands} & {\footnotesize  128 subbands} \\
& {\footnotesize  $\Delta$ToA= 1.175 $\mu$s} & {\footnotesize  $\Delta$ToA= 1.163 $\mu$s} & {\footnotesize  $\Delta$ToA= 1.16 $\mu$s} & {\footnotesize  $\Delta$ToA= 1.159 $\mu$s} & {\footnotesize  $\Delta$ToA= 1.159 $\mu$s} & {\footnotesize  $\Delta$ToA= 1.159 $\mu$s} \\
\hline                      
   {\footnotesize  1} & {\footnotesize  340007.11} & {\footnotesize  104292.40} & {\footnotesize  30135.74} & {\footnotesize  9825.05} & {\footnotesize  5252.04} & {\footnotesize  10598.98} \\    % inserting body of the table
   {\footnotesize $\left[1-2\right]$} & {\footnotesize  335893.74} & {\footnotesize  103686.56} & {\footnotesize  30077.98} & {\footnotesize  9822.81} & {\footnotesize  5255.34} & {\footnotesize  10603.07} \\
   {\footnotesize  $\left[1-3\right]$} & {\footnotesize  303811.64} & {\footnotesize  101127.57} & {\footnotesize  29947.06} & {\footnotesize  9829.48} & {\footnotesize  5262.52} & {\footnotesize  10611.12} \\
   {\footnotesize  $\left[1-4\right]$} & {\footnotesize  299319.89} & {\footnotesize  100419.18} & {\footnotesize  29893.99} & {\footnotesize  9833.71} & {\footnotesize  5270.98} & {\footnotesize  10620.33} \\
   {\footnotesize  $\left[1-5\right]$} & {\footnotesize  \textbf{275040.94}} & {\footnotesize  \textbf{96769.75}} & {\footnotesize  \textbf{29656.06}} & {\footnotesize  9834.93} & {\footnotesize  5279.79} & {\footnotesize  10629.55} \\
   {\footnotesize  $\left[1,3\right]$} & {\footnotesize  334793.14} & {\footnotesize  103567.04} & {\footnotesize  30068.92} & {\footnotesize  9822.36} & {\footnotesize  5255.59} & {\footnotesize  10603.31} \\
   {\footnotesize  $\left[1,3,5\right]$} & {\footnotesize  301516.21} & {\footnotesize  100859.88} & {\footnotesize  29928.58} & {\footnotesize  9828.86} & {\footnotesize  5262.68} & {\footnotesize  10611.16} \\
   {\footnotesize  2} & {\footnotesize  338727.76} & {\footnotesize  104150.09} & {\footnotesize  30122.78} & {\footnotesize  9823.52} & {\footnotesize  5251.60} & {\footnotesize  10598.57} \\
   {\footnotesize  $\left[2,4\right]$} & {\footnotesize  332738.84} & {\footnotesize  103339.95} & {\footnotesize  30052.10} & {\footnotesize  \textbf{9821.75}} & {\footnotesize  5256.13} & {\footnotesize  10603.84} \\
   {\footnotesize  No FD} & {\footnotesize  362504.68} & {\footnotesize  106036.77} & {\footnotesize  30258.10} & {\footnotesize  9827.26} & {\footnotesize  \textbf{5245.25}} & {\footnotesize  \textbf{10591.27}} \\
\hline                                  
\end{tabularx}
\end{table*}

\subsubsection{Subband optimisation from F-statistics}
\label{sssec:subband_optimisation_fcn_fstatistic}
\begin{figure}[!htbp]
\centering
\includegraphics[width=0.49\textwidth,clip]{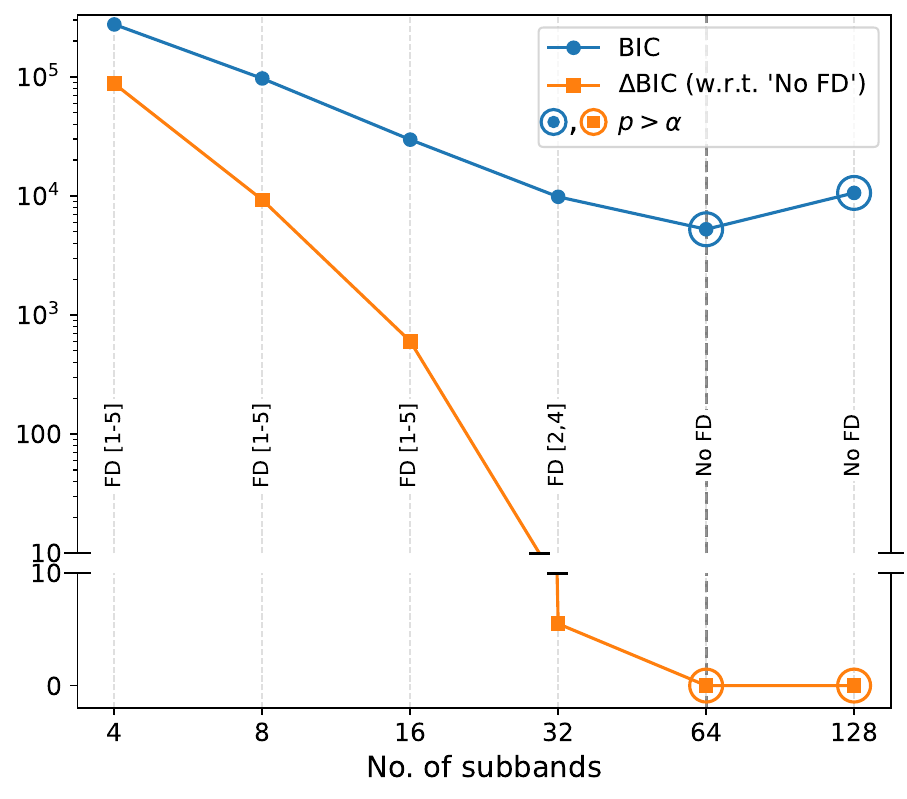}
\caption{BIC and $\Delta$BIC (w.r.t `No FD' model) of the preferred FD models obtained with FC$_n$ templates are plotted against each subband. The vertical grid lines correspond to the selected FD models. A thick dashed line represents the optimal number of subbands (64) where no FD parameter seems to be necessary. The circled points indicate p-values $> \alpha$ (0.05), implying that reductions in $\chi^2$ due to FD fits are statistically insignificant. The vertical axis is split into logarithmic and linear scales at BIC (or $\Delta BIC$) $=10$ for convenience of visualisation. }
\label{fig:bic_p_fcn}
\end{figure}

\begin{figure*}[!htbp]
\centering
\begin{tabularx}{\textwidth} { 
  >{\centering\arraybackslash}X
  >{\centering\arraybackslash}X } 
\includegraphics[width=0.4\textwidth,clip,trim=0.8cm 0.6cm 4.5cm 1.7cm]{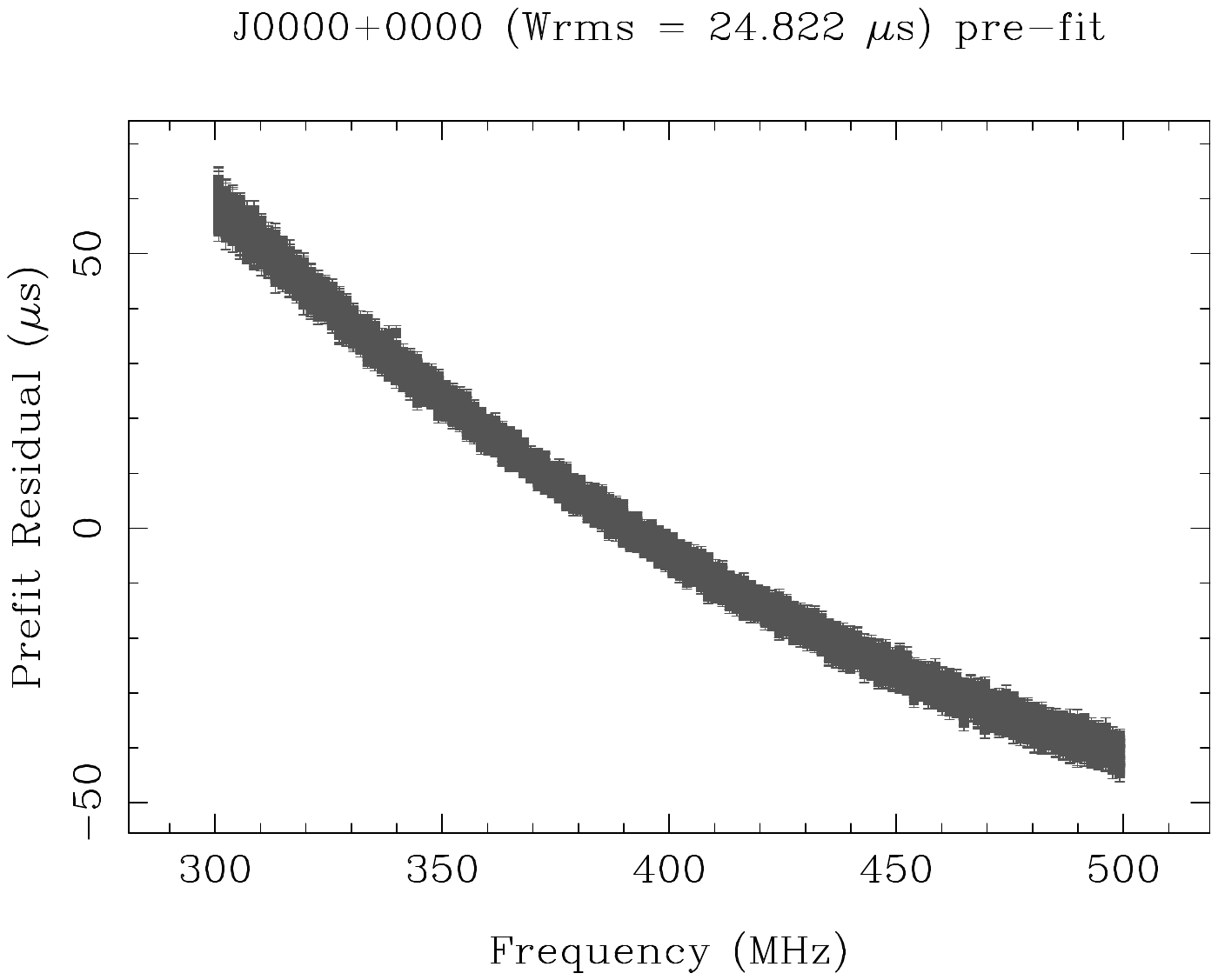} & \includegraphics[width=0.4\textwidth,clip,trim=0.8cm 0.6cm 4.5cm 1.7cm]{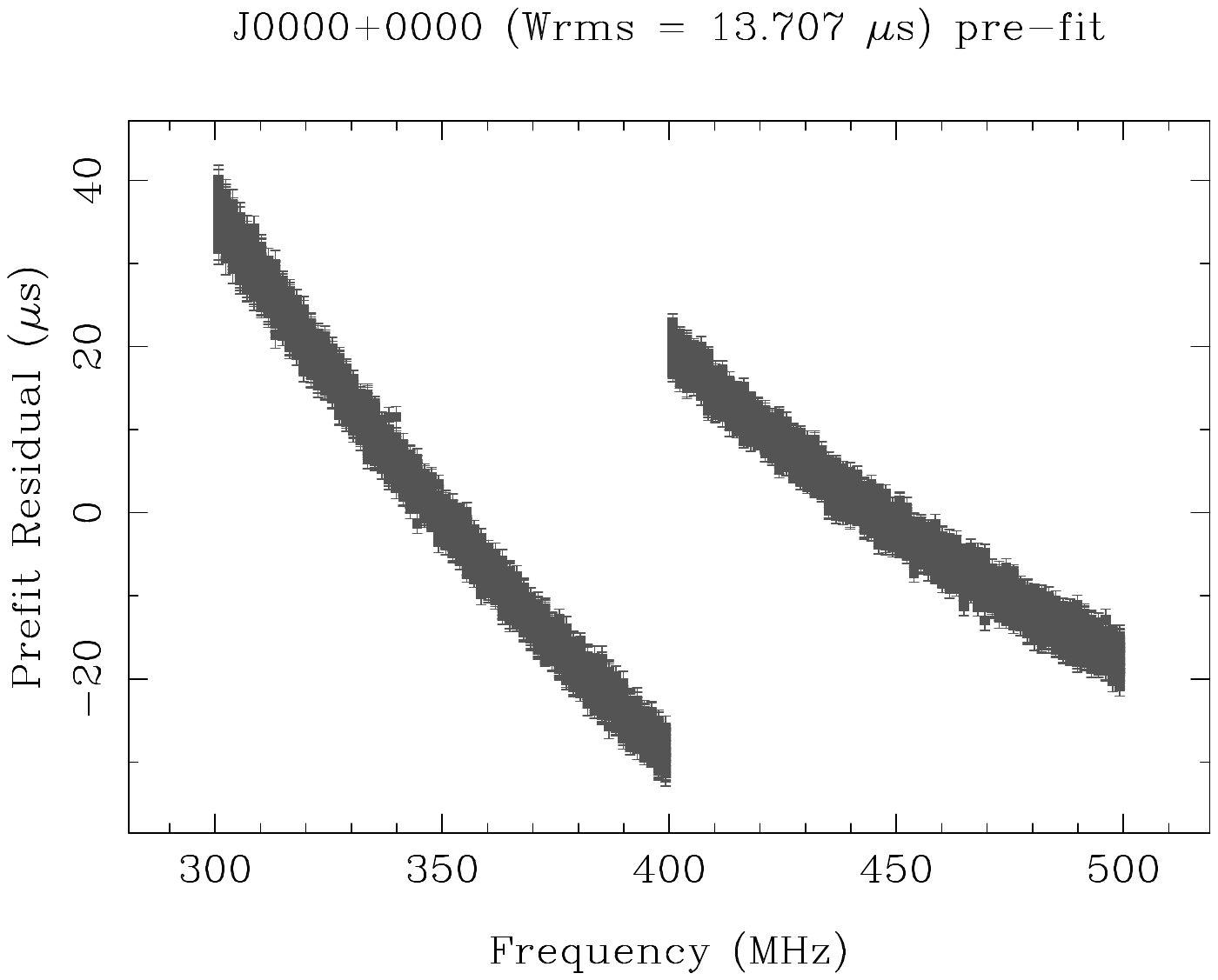} \\ \includegraphics[width=0.4\textwidth,clip,trim=0.8cm 0.6cm 4.5cm 1.7cm]{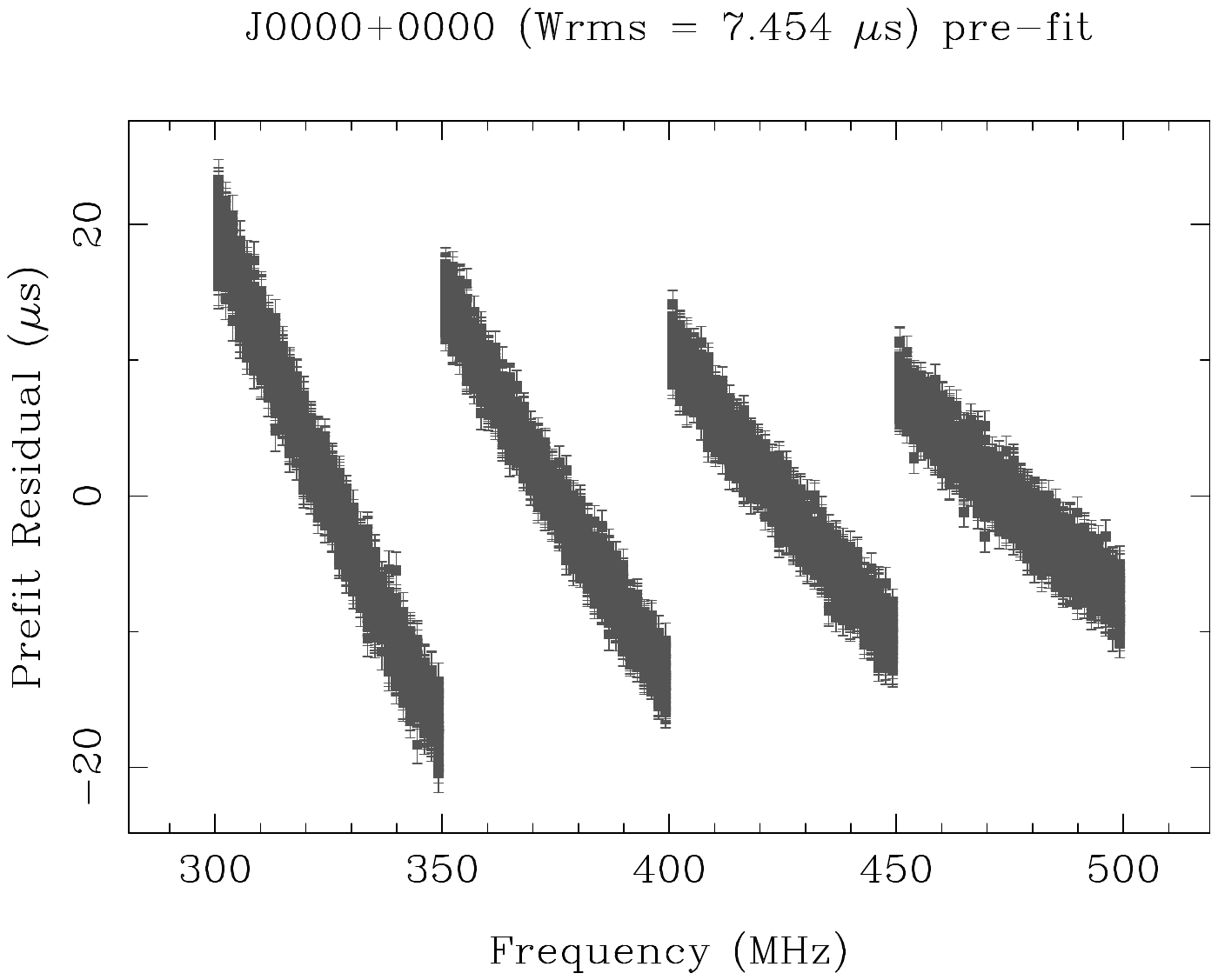} & \includegraphics[width=0.4\textwidth,clip,trim=0.8cm 0.6cm 4.5cm 1.7cm]{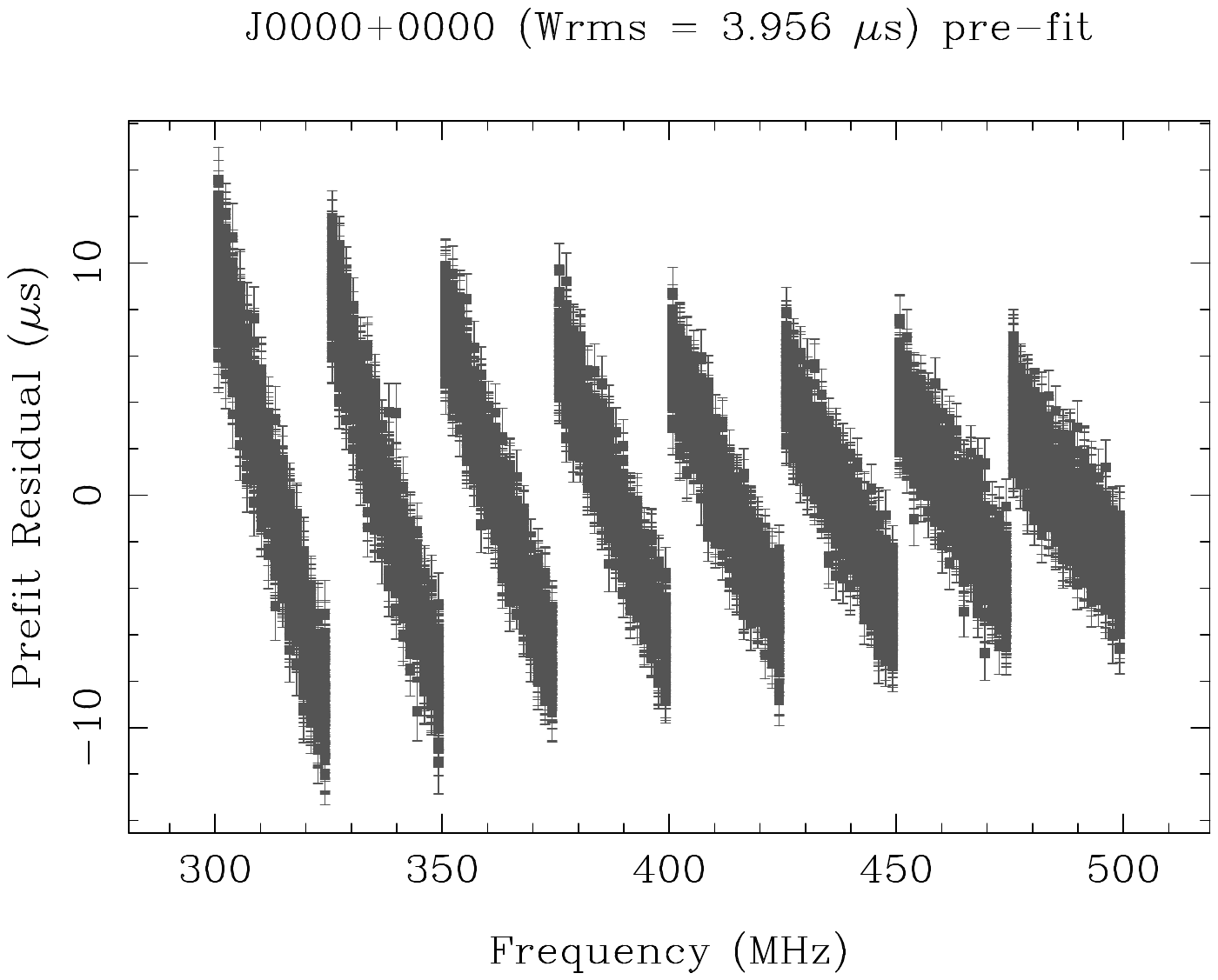} \\
\includegraphics[width=0.4\textwidth,clip,trim=0.8cm 0.6cm 4.5cm 1.7cm]{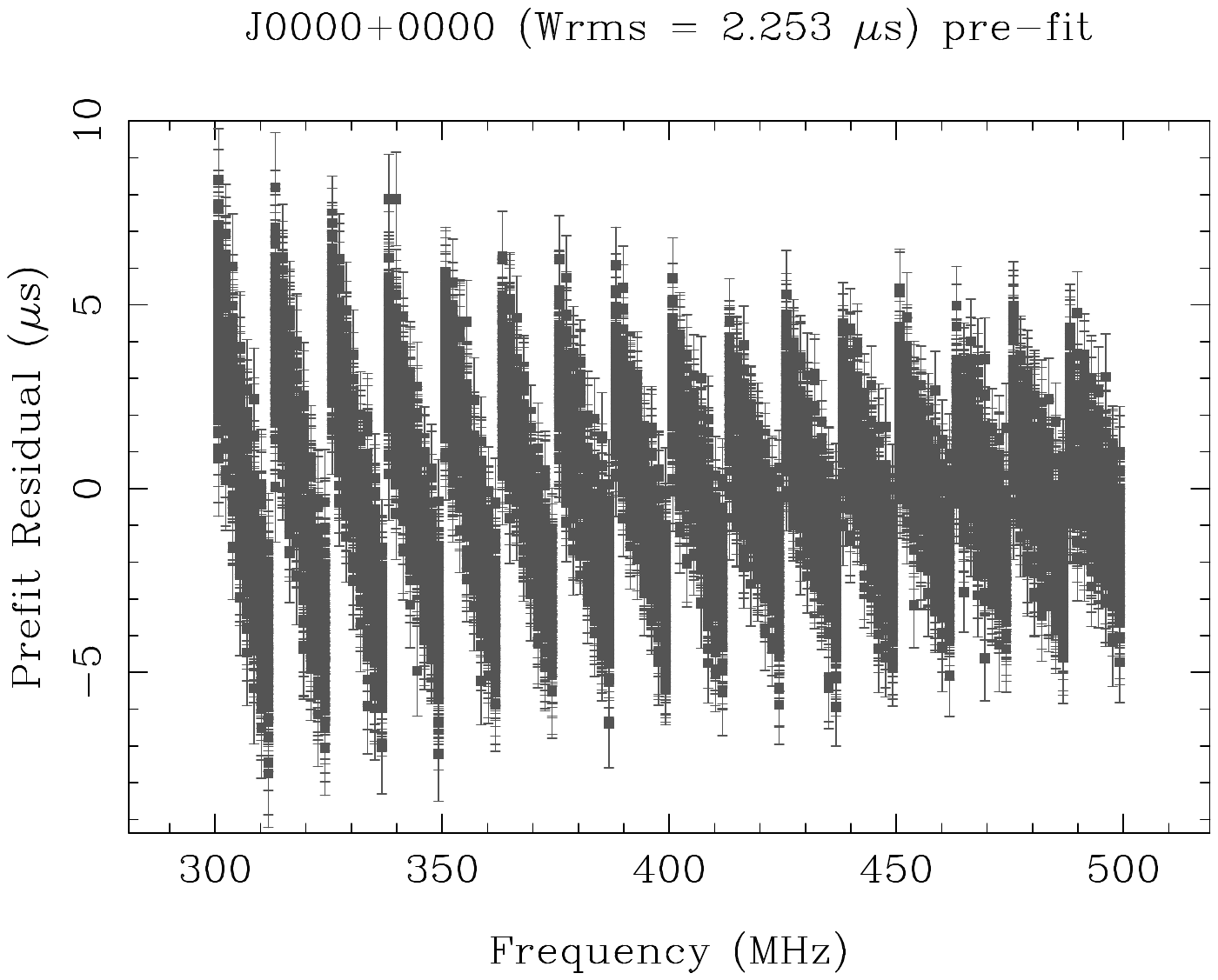} & \includegraphics[width=0.4\textwidth,clip,trim=0.8cm 0.6cm 4.5cm 1.7cm]{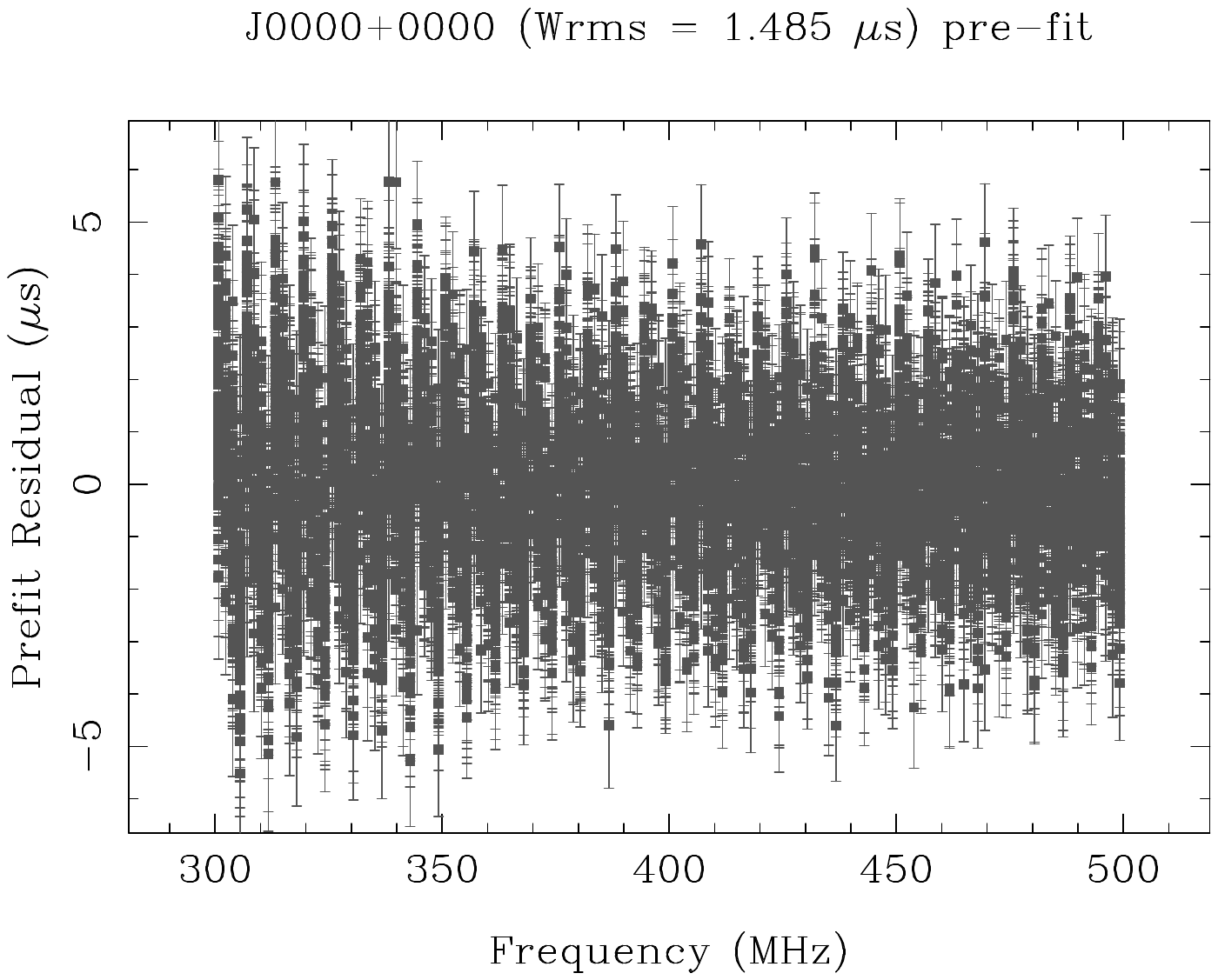} \\ \includegraphics[width=0.4\textwidth,clip,trim=0.8cm 0.6cm 4.5cm 1.7cm]{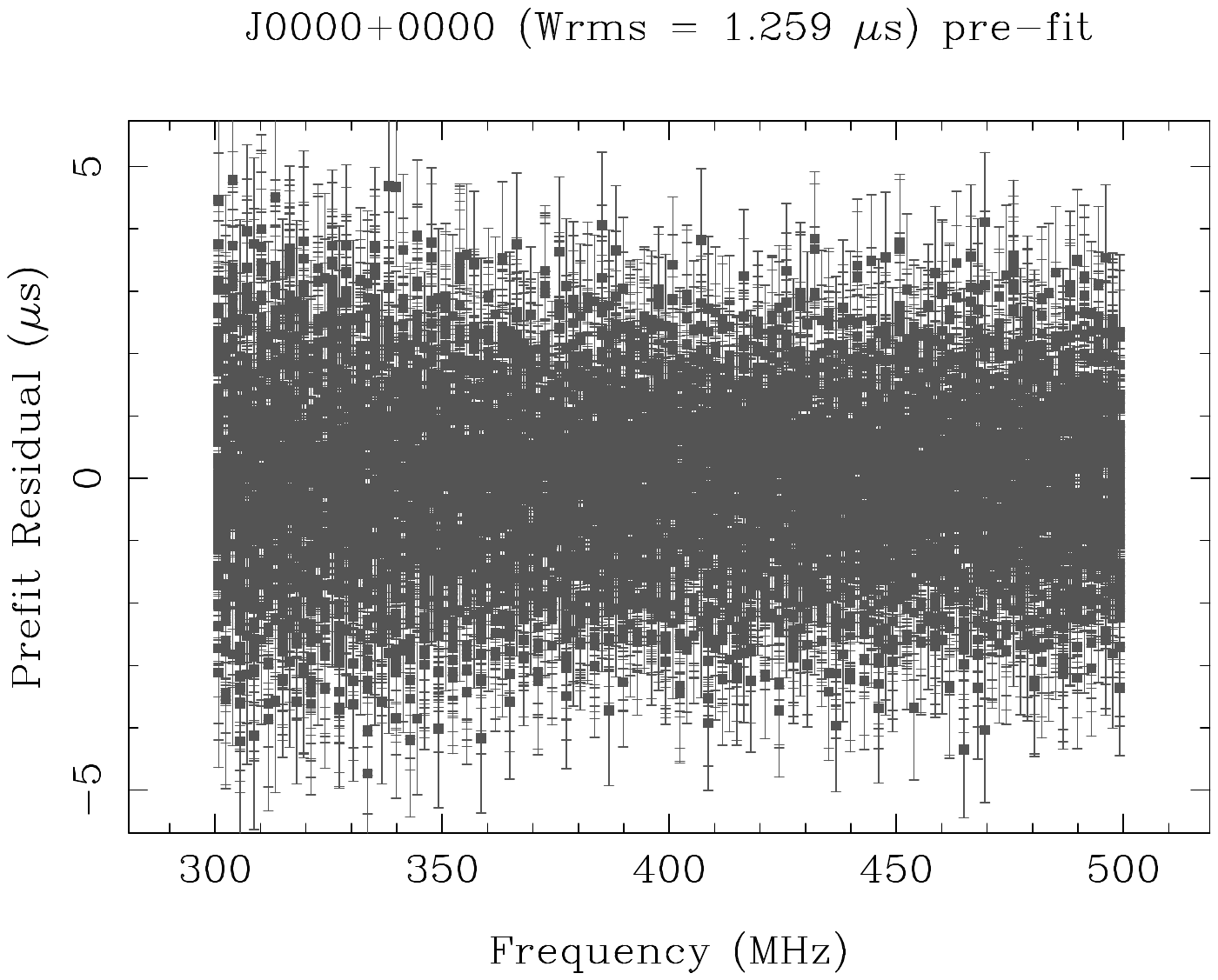} & \includegraphics[width=0.4\textwidth,clip,trim=0.8cm 0.6cm 4.5cm 1.7cm]{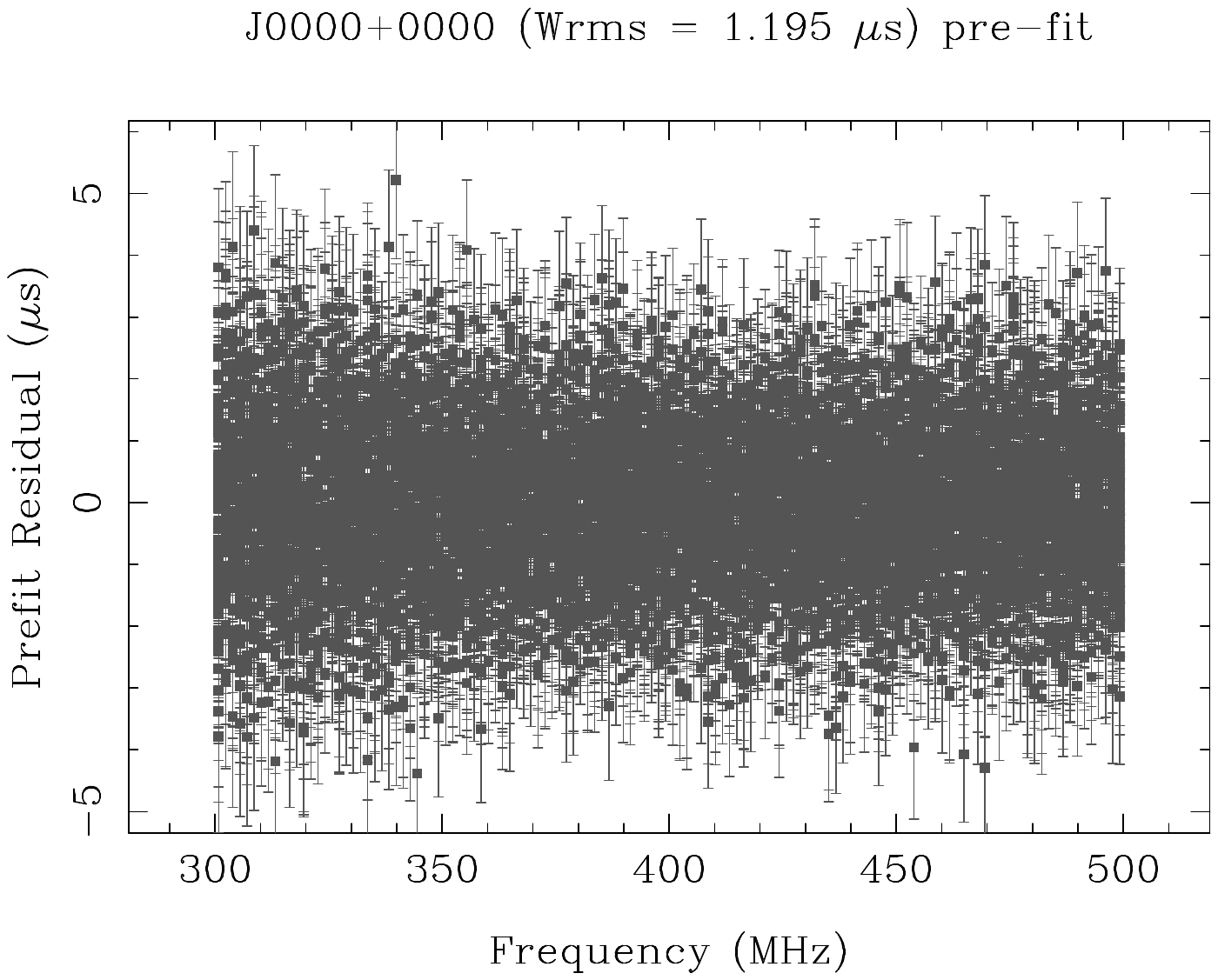} \\
\end{tabularx}
\caption{ToA residuals obtained using FC$_n$ templates vs. frequency for number of subbands increasing from 1 to 128 in steps of $n=2^k$ where $k \in \{0,1,2,3,4,5,6,7\}$. Top panel: $k=0$ (left), $k=1$ (right); second panel from top: $k=2$ (left), $k=3$ (right); third panel from top: $k=4$ (left), $k=5$ (right); bottom panel: $k=6$ (left), $k=7$ (right).}
\label{fig:residuals_fcn}
\end{figure*}

\begin{figure*}[!htbp]
\centering
\begin{tabularx}{\textwidth} { 
  >{\centering\arraybackslash}X
  >{\centering\arraybackslash}X } 
\includegraphics[width=0.45\textwidth,clip,trim=0.6cm 0.5cm 2.5cm 1.5cm]{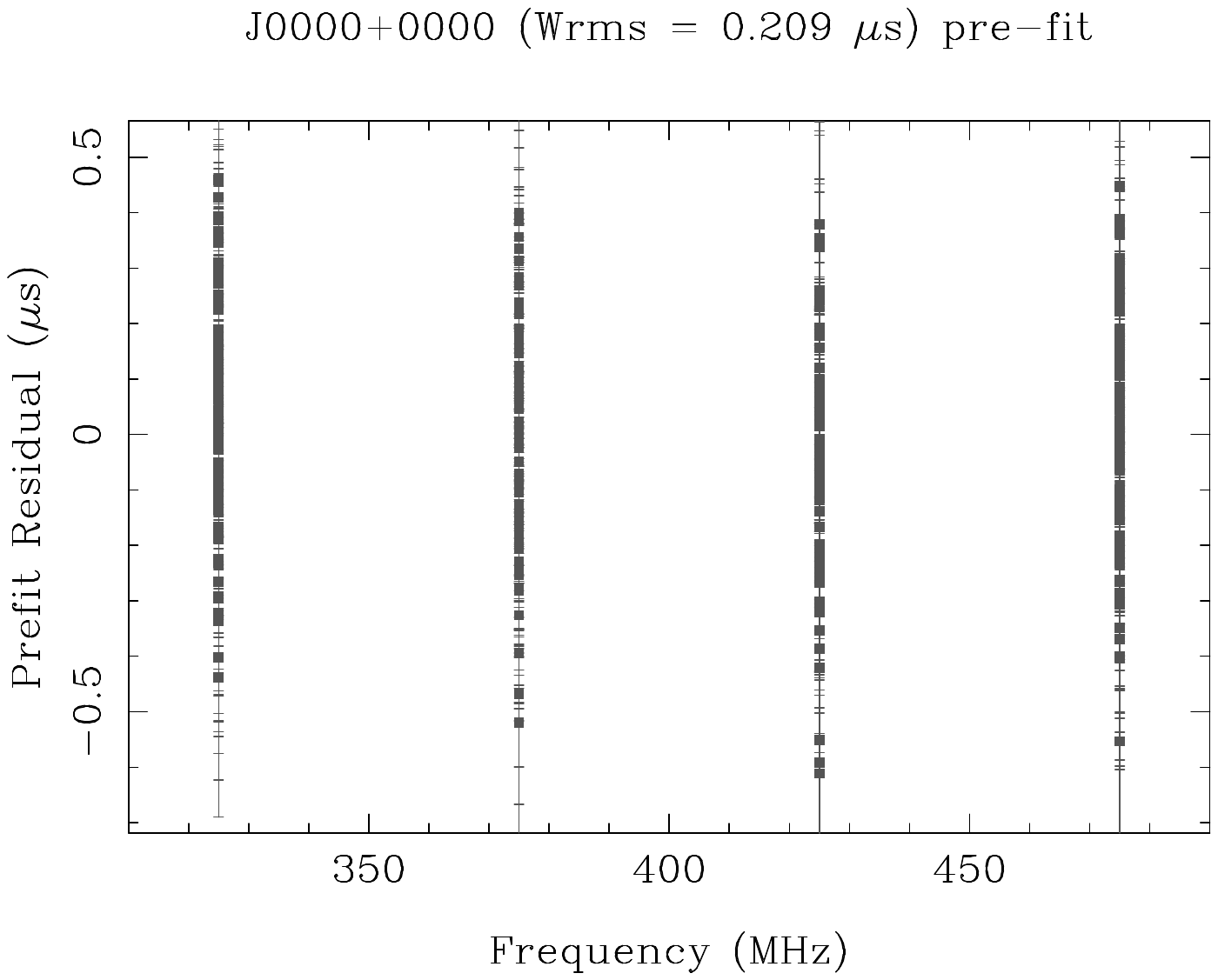} & \includegraphics[width=0.45\textwidth,clip,trim=0.6cm 0.5cm 2.5cm 1.5cm]{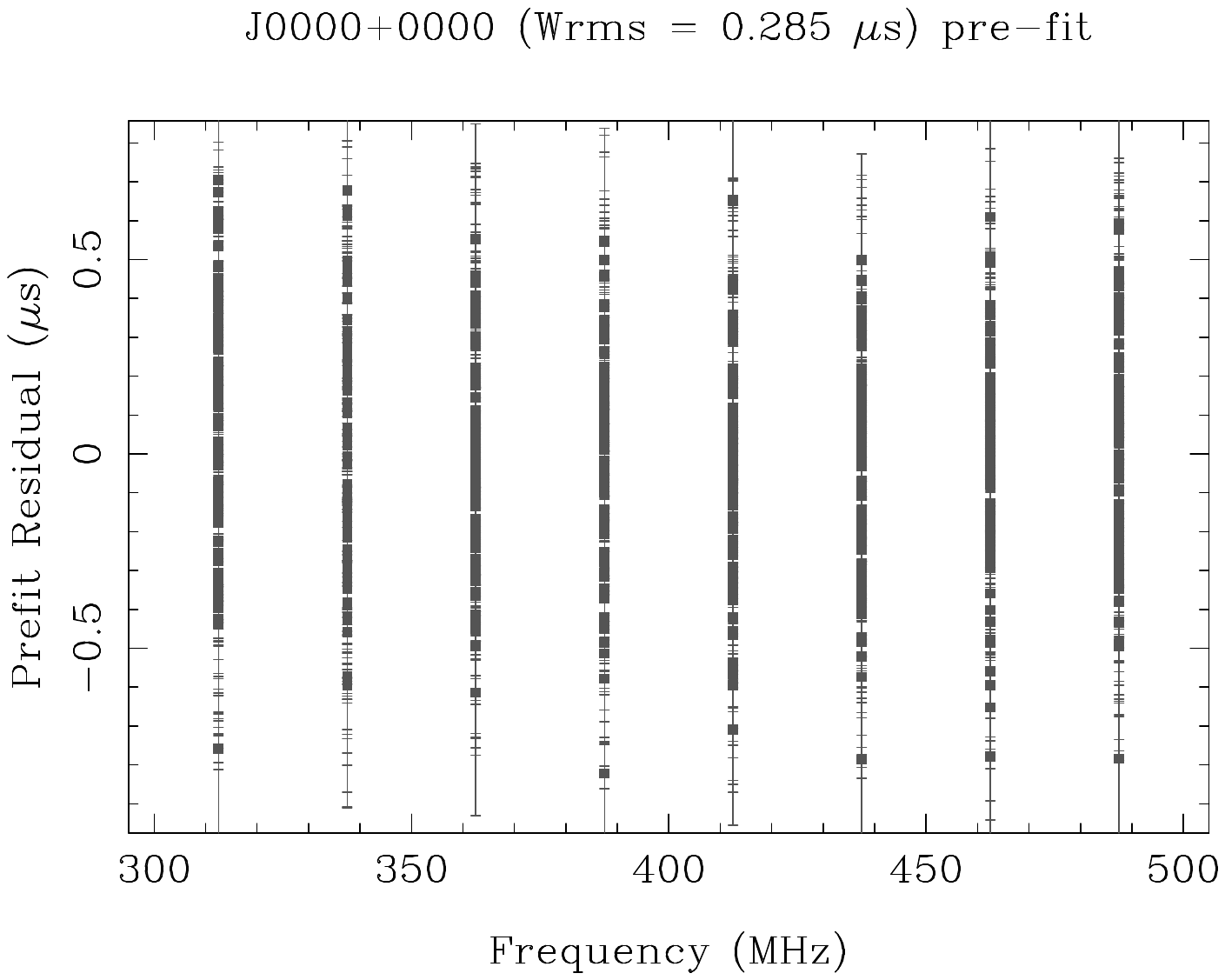} \\ \includegraphics[width=0.45\textwidth,clip,trim=0.6cm 0.5cm 2.5cm 1.5cm]{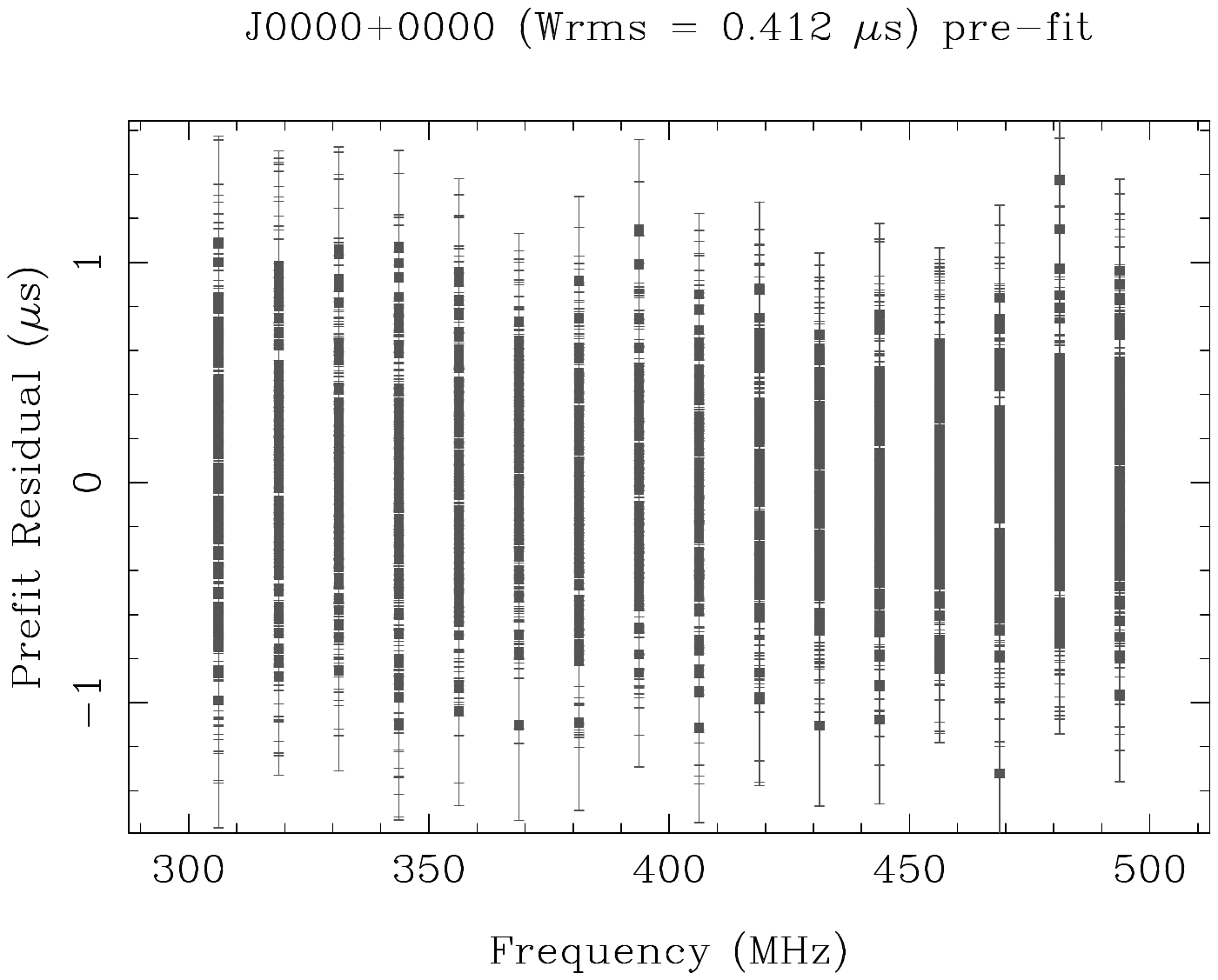} &
\includegraphics[width=0.45\textwidth,clip,trim=0.6cm 0.5cm 2.5cm 1.5cm]{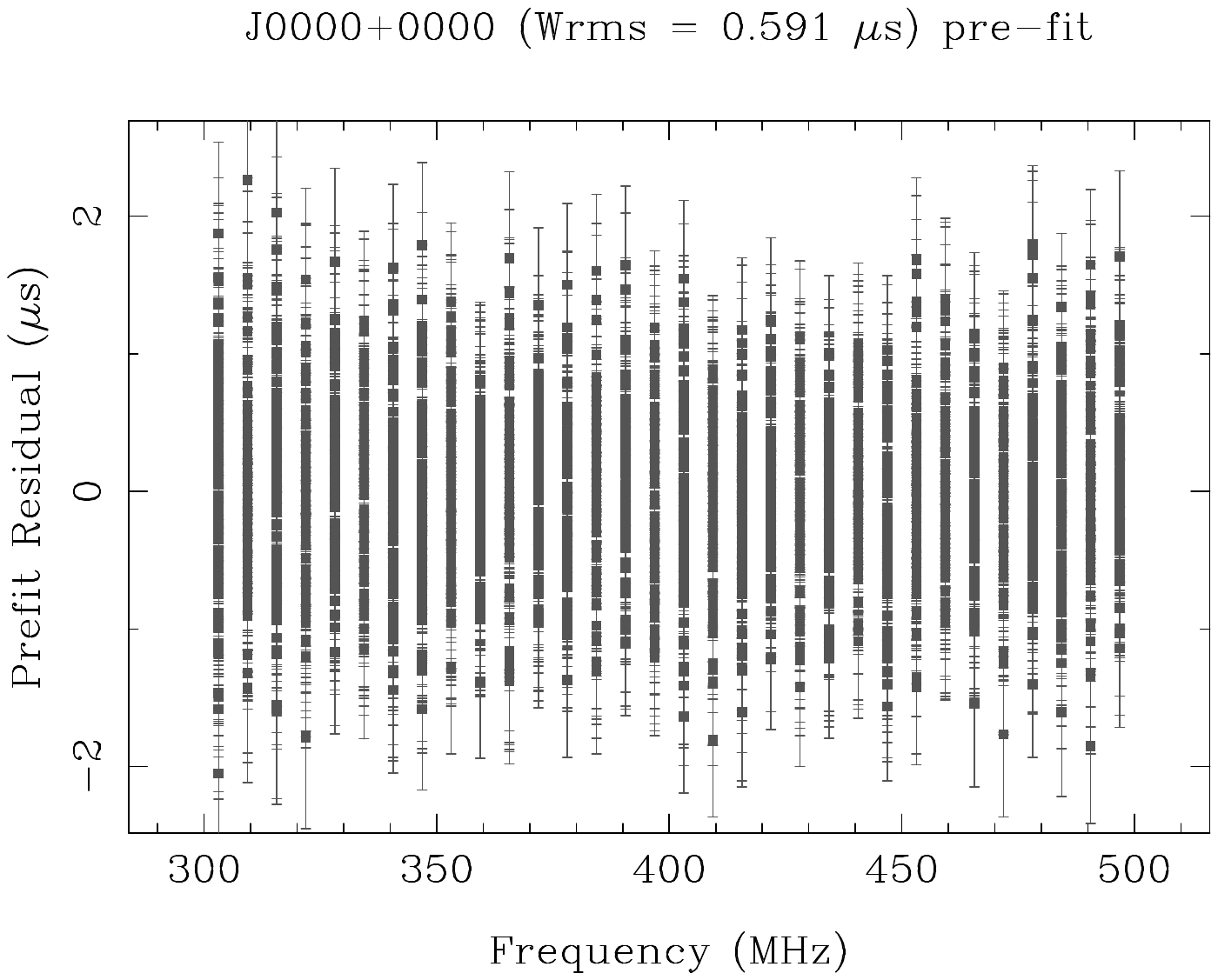} \\
\includegraphics[width=0.45\textwidth,clip,trim=0.6cm 0.5cm 2.5cm 1.5cm]{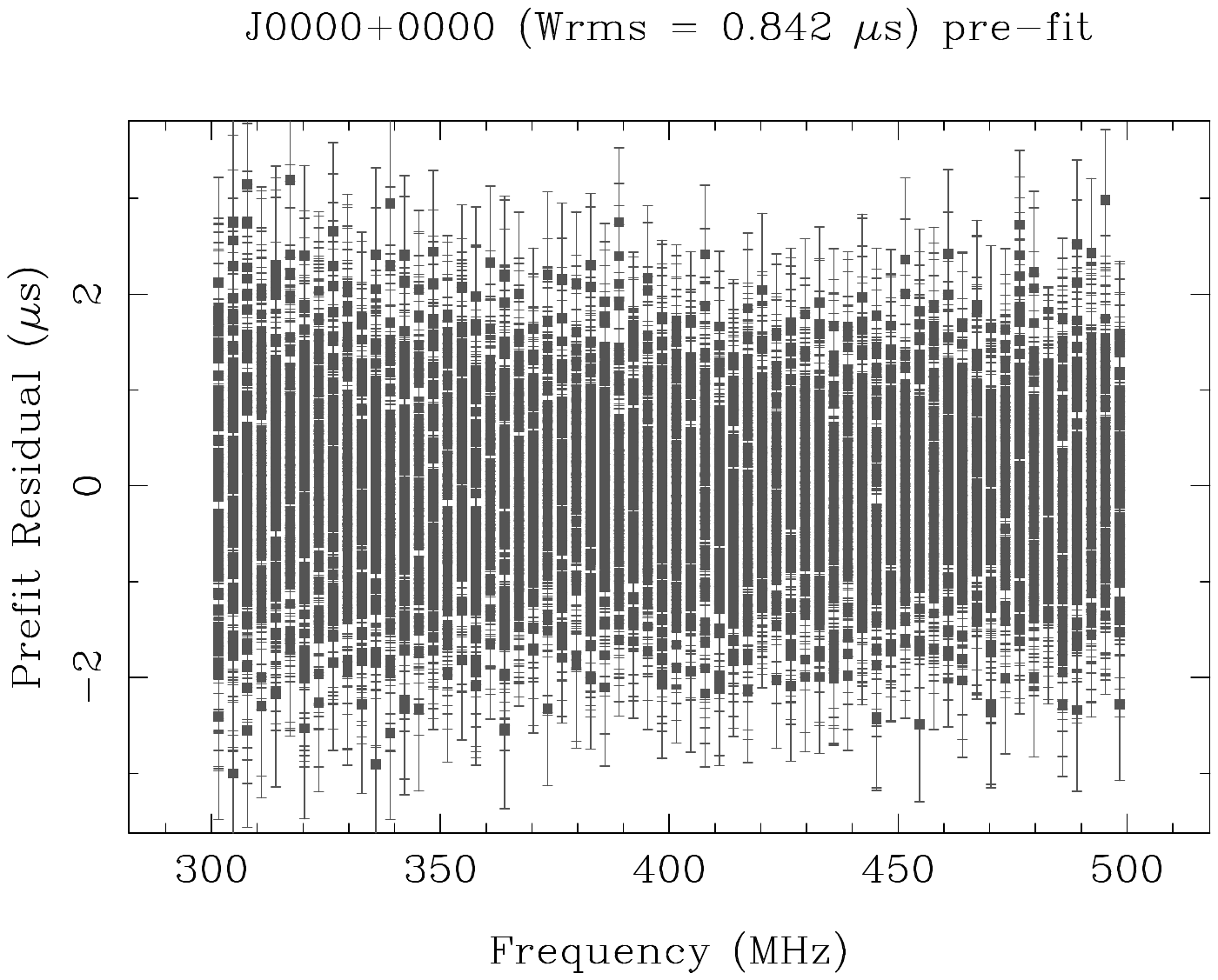} & \includegraphics[width=0.45\textwidth,clip,trim=0.6cm 0.5cm 2.5cm 1.5cm]{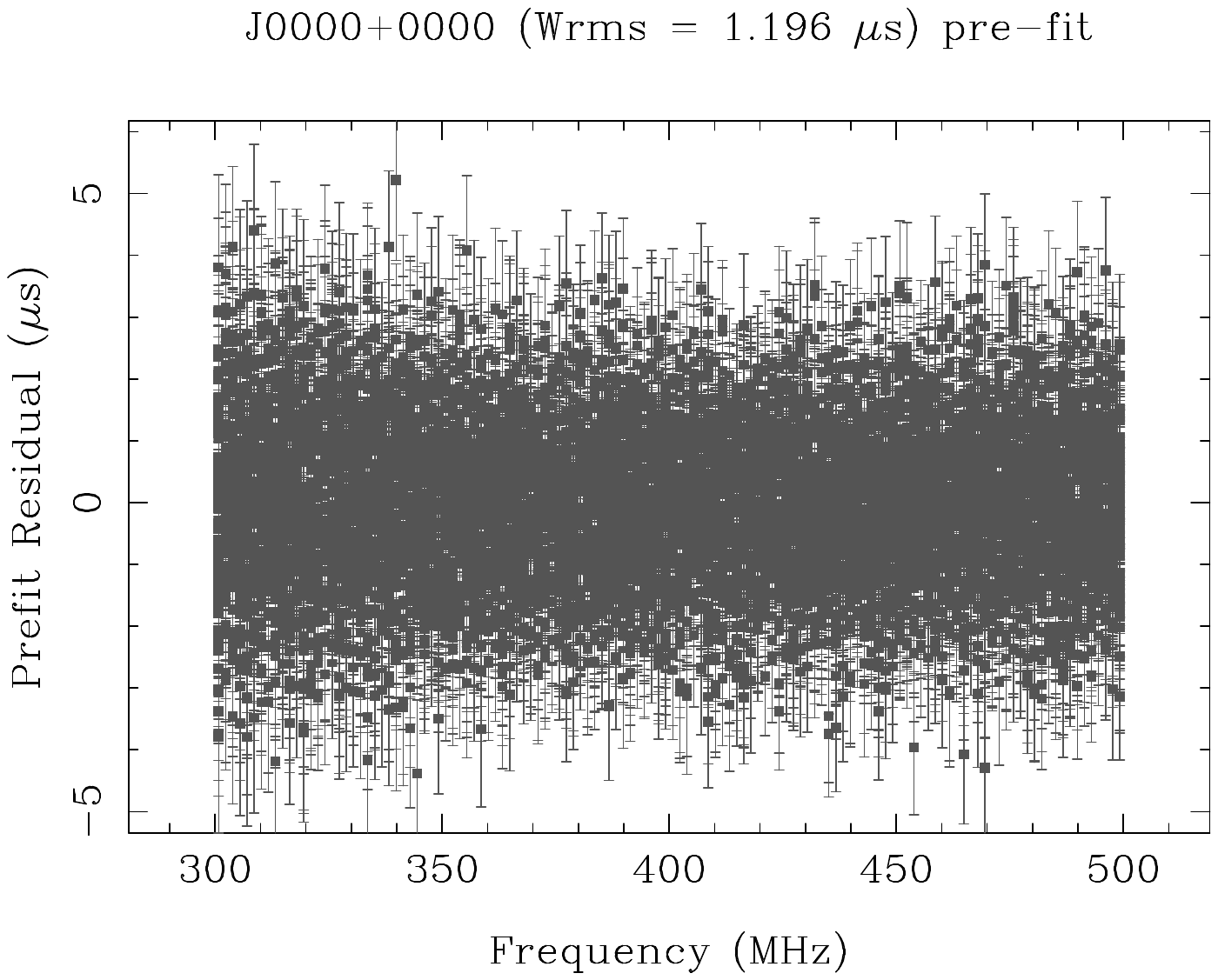} \\ 
\end{tabularx}
\caption{Subbanded ToA residuals obtained using FR templates vs. frequency (4, 8, 16, 32, 64 \& 128 subbands)}
\label{fig:residuals_fr}
\end{figure*}

\noindent
The F-ratios and the corresponding p-values for all possible configurations listed in table \ref{tab:fd_model_select_fcn} are calculated from equation \ref{eqn:f-value} and F$_c$ for a given significance $\alpha$ as explained in section \ref{sssec:subband_optimisation_fr_fstatistic}. The p-values are presented in table \ref{tab:subband_opt_fcn}. All p-values below significance $\alpha$ ($0.05$) are marked in boldface. We note that the turn-over from p $<\alpha$ to p $>\alpha$ occurs between 32 and 64 subbands implying \textit{a transition from FD-significant to FD-indifferent} timing model. The transition can be clearly seen in effect at 32 subbands where majority of FD combinations appear significant (including the best model as per BIC) except two models with a single FD parameter (FD1/FD2) whose p-values are above $\alpha$. This is in perfect agreement with the results of FD model selection from BIC as presented in table \ref{tab:fd_model_select_fcn}, and in contrast to the results for ToAs obtained from FR templates listed in table \ref{tab:subband_opt_fr} where no such transition is noticed. While the latter suggests no subband optimisation and prefers a `No FD' model at any number of subbands permitted by the available frequency resolution, the former clearly shows the necessity of FD parameters below a given number of subbands. The optimal number will certainly depend on the steepness of profile evolution and hence is expected to be pulsar-specific. A graphical summary of the results presented in tables \ref{tab:fd_model_select_fcn} and \ref{tab:subband_opt_fcn} is depicted in figure \ref{fig:bic_p_fcn}. Also, a visual demonstration of the transition from FD to `No FD' model with progressively increasing number of subbands is depicted in figure \ref{fig:residuals_fcn} where ToA residuals obtained using FC$_n$ templates are plotted against frequency. Though the selection of FD models and the subband optimisation have been done based on comparision of fit statistics 4-subband onwards for reasons stated in section \ref{sssec:model_selection_fcn_bic}, ToA residuals from FC$_1$ (frequency-averaged template across full 200 MHz) and FC$_2$ (2 frequency-averaged templates, each 100 MHz wide) are also shown in the plot to demonstrate a gradual diminution of the FD trend until such subband-resolution is achieved where frequency-dependence becomes practically invisible to be mitigated with FD parameters. A similar plot depicting subbanded ToA residuals (4, 8, 16, 32, 64 \& 128) from FR template is shown in figure \ref{fig:residuals_fr} for comparision, where no FD trends are visible unlike figure \ref{fig:residuals_fcn}. In the following section, we further attempt to consolidate the proposed optimisation from statistical significance by doing a direct comparision with any visible residual pulse between adjacent subbands in the profile domain.  

\begin{table*}[ht!]
\caption{Subband optimisation using p-values from F-statistic (FC$_n$ templates). Boldfaces represent p $<\alpha$ (0.05). $\Delta$ToAs are medians of ToA uncertainties across a given subband.}                 
\label{tab:subband_opt_fcn}    
\centering     
\begin{tabularx}{\textwidth} { 
  >{\centering\arraybackslash}X
  >{\centering\arraybackslash}X
  >{\centering\arraybackslash}X
  >{\centering\arraybackslash}X
  >{\centering\arraybackslash}X
  >{\centering\arraybackslash}X
  >{\centering\arraybackslash}X }      
\hline\hline               
{\footnotesize FD Model} & {\footnotesize 4 subbands} & {\footnotesize 8 subbands} & {\footnotesize 16 subbands} & {\footnotesize 32 subbands} & {\footnotesize 64 subbands} & {\footnotesize 128 subbands} \\
& {\footnotesize $\Delta$ToA= 1.175 $\mu$s} & {\footnotesize $\Delta$ToA= 1.163 $\mu$s} & {\footnotesize $\Delta$ToA= 1.16 $\mu$s} & {\footnotesize $\Delta$ToA= 1.159 $\mu$s} & {\footnotesize $\Delta$ToA= 1.159 $\mu$s} & {\footnotesize $\Delta$ToA= 1.159 $\mu$s} \\
\hline                      
   {\footnotesize 1} & {\footnotesize \textbf{1.9$\mathbf{\times 10^{-39}}$}} & {\footnotesize \textbf{3.1$\mathbf{\times 10^{-11}}$}} & {\footnotesize \textbf{0.0007}} & {\footnotesize 0.0896} & {\footnotesize 0.3497} & {\footnotesize 0.5400} \\    % inserting body of the table
   {\footnotesize$\left[1-2\right]$} & {\footnotesize \textbf{4.8$\mathbf{\times 10^{-58}}$}} & {\footnotesize \textbf{2.2$\mathbf{\times 10^{-17}}$}} & {\footnotesize \textbf{1.6$\mathbf{\times 10^{-5}}$}} & {\footnotesize \textbf{0.0247}} & {\footnotesize 0.1057} & {\footnotesize 0.3468} \\
   {\footnotesize $\left[1-3\right]$} & {\footnotesize \textbf{1.8$\mathbf{\times 10^{-156}}$}} & {\footnotesize \textbf{2.0$\mathbf{\times 10^{-40}}$}} & {\footnotesize \textbf{1.9$\mathbf{\times 10^{-9}}$}} & {\footnotesize \textbf{0.0262}} & {\footnotesize 0.1039} & {\footnotesize 0.4124} \\
   {\footnotesize $\left[1-4\right]$} & {\footnotesize \textbf{2.8$\mathbf{\times 10^{-185}}$}} & {\footnotesize \textbf{4.0$\mathbf{\times 10^{-50}}$}} & {\footnotesize \textbf{1.5$\mathbf{\times 10^{-11}}$}} & {\footnotesize \textbf{0.0171}} & {\footnotesize 0.1454} & {\footnotesize 0.5324} \\
   {\footnotesize $\left[1-5\right]$} & {\footnotesize \textbf{3.1$\mathbf{\times 10^{-291}}$}} & {\footnotesize \textbf{1.0$\mathbf{\times 10^{-90}}$}} & {\footnotesize \textbf{9.1$\mathbf{\times 10^{-20}}$}} & {\footnotesize \textbf{0.0060}} & {\footnotesize 0.2157} & {\footnotesize 0.6419} \\
   {\footnotesize $\left[1,3\right]$} & {\footnotesize \textbf{1.4$\mathbf{\times 10^{-60}}$}} & {\footnotesize \textbf{3.0$\mathbf{\times 10^{-18}}$}} & {\footnotesize \textbf{9.5$\mathbf{\times 10^{-6}}$}} & {\footnotesize \textbf{0.0228}} & {\footnotesize 0.1145} & {\footnotesize 0.3603} \\
   {\footnotesize $\left[1,3,5\right]$} & {\footnotesize \textbf{1.4$\mathbf{\times 10^{-163}}$}} & {\footnotesize \textbf{1.0$\mathbf{\times 10^{-42}}$}} & {\footnotesize \textbf{5.7$\mathbf{\times 10^{-10}}$}} & {\footnotesize \textbf{0.0234}} & {\footnotesize 0.1094} & {\footnotesize 0.4147} \\
   {\footnotesize 2} & {\footnotesize \textbf{1.0$\mathbf{\times 10^{-41}}$}} & {\footnotesize \textbf{4.8$\mathbf{\times 10^{-12}}$}} & {\footnotesize \textbf{0.0004}} & {\footnotesize 0.0696} & {\footnotesize 0.2952} & {\footnotesize 0.4896} \\
   {\footnotesize $\left[2,4\right]$} & {\footnotesize \textbf{2.4$\mathbf{\times 10^{-65}}$}} & {\footnotesize \textbf{6.9$\mathbf{\times 10^{-20}}$}} & {\footnotesize \textbf{3.7$\mathbf{\times 10^{-6}}$}} & {\footnotesize \textbf{0.0206}} & {\footnotesize 0.1364} & {\footnotesize 0.3921} \\
\hline                                  
\end{tabularx}
\end{table*} 

%%%%%%%%%%%%%%%%%%%%%%%%%%%%%%%%%%%%%%%%%%%%%%%%%%%%%%%%%%%%%%
\section{Subband optimisation in profile domain}
\label{sec:profile_domain} 
\noindent
In this section, we perform a visual inspection of profile residuals across adjacent subbands to determine the number of subbands required for mitigating intrinsic FD-effects. The test is to arrive at an optimal number and compare it with that obtained from the statistical inferences elaborated above. In principle, this method does not depend on the specifics of the template generation technique or ToA estimation algorithm. It provides a direct estimate of the minimum subbanding needed to eliminate FD trends in timing residuals related to intrinsic pulse profile evolution. 

\begin{figure*}[!htbp]
\centering
\includegraphics[width=\textwidth,clip,trim=2.7cm 2.0cm 2.5cm 0cm]{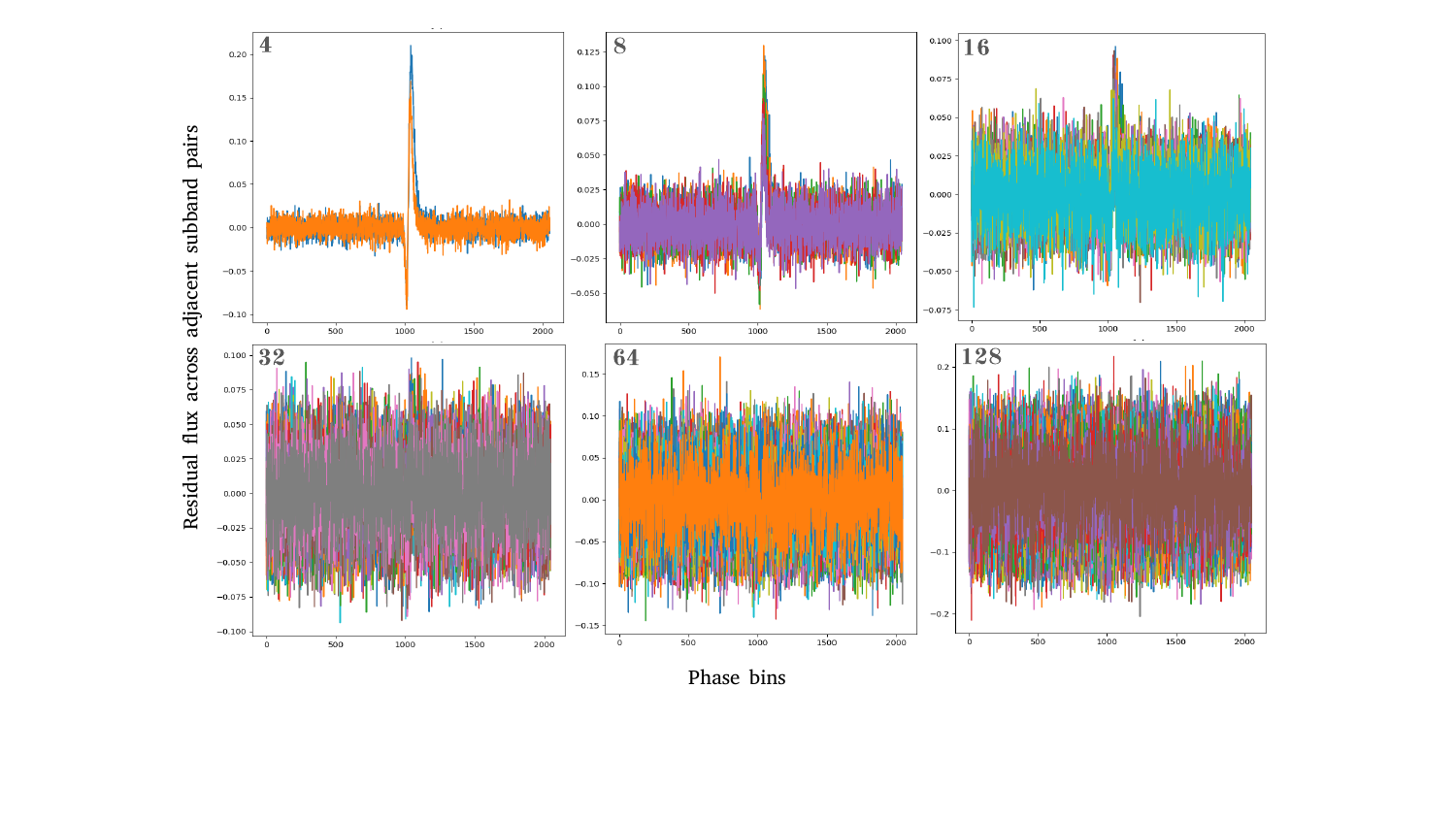}
\caption{Residual flux across adjacent subbands for $4$, $8$, $16$, $32$, $64$ and $128$ subbanded configurations plotted against the number of phase bins. Different colours correspond to different adjacent subband pairs.}
\label{fig:subband_selection}
\end{figure*}

Figure \ref{fig:subband_selection} depicts the residual profiles across adjacent subbands plotted against phase bins for $4$, $8$, $16$, $32$, $64$ and $128$ subbanded configurations respectively. The difference profiles are normalised with their peak values for a clear visual comparision. It is evident that, for the given simulated pulse, any remnant signal across adjacent subbands due to profile frequency evolution is masked by the noise floor beyond $32$ subbands. The observation matches exactly with the optimum number suggested by BIC and F-statistics from the comparision of fitted models for ToAs estimated with FC$_n$ templates. It also agrees with the gradual mitigation of the visible FD trend with an increasing number of subbands in the timing residuals obtained from FC$_n$ templates as depicted in figure \ref{fig:residuals_fcn}. The comparision of difference profiles between adjacent subbands thus provides validation and a one-to-one correspondence between frequency (timing) and profile domains, converging coherently to the same outcome.  

%%%%%%%%%%%%%%%
\section{Concluding remarks}
\label{sec:conclusion}
\noindent
Our analysis verifies that FD parameters are required to mitigate chromatic delays in timing residuals estimated from ToAs generated by cross-correlation with templates that are frequency-averaged over either the entire band or partial segments of the bandwidth. Our results show that a judicious optimisation in subbanding of the frequency-averaged templates for ToA generation can do away with the necessity of FD parameters in timing solutions. The optimisation will depend on the steepness and overall nature of profile-frequency evolution, and hence, will be pulsar-specific. It is of particular importance for PTA datasets where subbanded ToAs are generated from templates collapsed across wide frequency bands or parts of it prior to cross-correlation with archives. Finally, our results establish within the given simulation framework, that the use of frequency-resolved templates for subbanded ToA generation eliminates the significance of FD parameters altogether. While FD fits to these residuals \textit{may} still decrease the weighted RMS by a few nanoseconds with marginal improvement in the reduced $\chi^2$, our investigation confirms -- based on BIC, F-statistics, and visual inspection of timing residuals and subbanded pulse profiles -- that such improvements are statistically inconsequential. Therefore, avoiding such fits is solicited for the sake of retaining physically relevant chromatic trends that, without careful and in-depth study, would otherwise get absorbed in the FD parameters. It is worth mentioning that for widely separated multi-band PTA observations, the optimisation of subbands needs to be achieved for each band in a way that the S/N per subband remains comparable across the entire bandwidth \citep{Rana25}. This requires striking a careful balance between the examination of fit statistics and that of the estimated ToA uncertainties. A related study on the significance of FD parameters for multi-component profile evolution is being pursued by \cite{DD26} (under review). We do not attempt noise modelling in the present work as our simulation incorporates profile evolution across frequency with a constant DM devoid of stochastic variations of the ISM. The injected Gaussian noise is white in nature, and any frequency-dependent delay in the timing residuals is attributed exclusively to what the FD parameters are meant to model. Simulation of long-term deterministic variations and short-term stochastic variations of the ISM in terms of both dispersion and scattering for comparision with real timing datasets is being investigated in our upcoming works.
%%%%%%%%%%%%%%%%%%%%%%%%%%%%%%%%%%%%%%%%%%%%%%%%%%%%%%%%%%%%%%
\section{Software}
\noindent
\texttt{Python}, \texttt{NumPy}, \texttt{SciPy}, \texttt{Matplotlib}, \texttt{PSRCHIVE}, \texttt{TEMPO2}, pulsar profile simulator being developed in-house as part of the InPTA experiment (in development for release).
%%%%%%%%%%%%%%%%%%%%%%%%%%%%%%%%%%%%%%%%%%%%%%%%%%%%%%%%%%%%%%
\section{Data availability}
\noindent
Data used in this work will be made available upon request.
%%%%%%%%%%%%%%%%%%%%%%%%%%%%%%%%%%%%%%%%%%%%%%%%%%%%%%%%%%%%%%
\begin{acknowledgements}
PT acknowledges financial support from the Istituto Nazionale di Astrofisica (INAF) 2023 Large Grant: “Gravitational Wave Detection using Pulsar Timing Arrays” (PI: Delphine Perrodin). PT also acknowledges Delphine Perrodin and Tinn Thongmeearkom for providing useful comments on the manuscript. DD acknowledges the ilifu cloud computing facility-www.ilifu.ac.za. HG acknowledges the support from SERB-DST, Govt. of India, via project code No. CRG/2022/009359. BCJ acknowledges the support from Raja Ramanna Chair fellowship of the Department of Atomic Energy, Government of India (RRC – Track I Grant 3/3401 Atomic Energy Research 00 004 Research and Development 27 02 31 1002//2/2023/RRC/R\&D-II/13886 and 1002/2/2023/RRC/R\&D-II/14369). SD acknowledges support from MTR\_2023\_000384. The work of CD at the Physical Research Laboratory (PRL) is supported by the Department of Space, Government of India. JS acknowledges the support from the University of Cape Town Vice Chancellor’s Future Leaders 2030 Awards programme and the South African Research Chairs Initiative of the Department of Science and Technology and the National Research Foundation. AKP is supported by CSIR fellowship Grant number 09/0079(15784)/2022-EMR-I. KT is partially supported by JSPS KAKENHI grant Nos. 24H01813, 25K21670, 26H00838 and 26K21724.
\end{acknowledgements}

\bibliographystyle{bibtex/general}
\bibliography{main}

@ARTICLE{Akaike78,
       author = {{Akaike}, H},
        title = "{A new look at the statistical model identification}",
      journal = {IEEE Transactions on Automatic Control},
         year = 1974,
        volume = {19(6)},
        pages = {716--723},
          doi = {10.1109/TAC.1974.1100705}
}

@ARTICLE{Schwarz78,
       author = {{Schwarz}, G},
        title = "{Estimating the Dimension of a Model}",
      journal = {Annals of Statistics},
         year = 1978,
        volume = {6(2)},
        pages = {461--464},
          doi = {10.1214/aos/1176344136}
}

@ARTICLE{Cordes78,
       author = {{Cordes}, J.~M.},
        title = "{Observational limits on the location of pulsar emission regions.}",
      journal = {\apj},
         year = 1978,
        month = jun,
       volume = {222},
        pages = {1006-1011},
          doi = {10.1086/156218},
}

@BOOK{Bevington03,
       author = {{Bevington}, Philip R. and {Robinson}, D. Keith},
        title = "{Data reduction and error analysis for the physical sciences}",
         year = 2003,
         publisher = {McGraw Hill}, 
         edition = {3rd}, 
         isbn = {0-07-247227-8}
}

@ARTICLE{HVM04,
       author = {{Hotan}, AW and {van Straten}, W and {Manchester}, RN},
        title = "{psrchive and psrfits: An Open Approach to Radio Pulsar Data Storage and Analysis}",
      journal = {\pasa},
         year = 2004,
        volume = {21(3)},
        pages = {302--309},
          doi = {10.1071/AS04022}
}

@ARTICLE{HEM06,
       author = {{Hobbs}, G and {Edwards}, RT and {Manchester}, RN},
        title = "{TEMPO2, a new pulsar-timing package – I. An overview}",
      journal = {\mnras},
         year = 2006,
        volume = {369},
        pages = {655--672},
          doi = {10.1111/j.1365-2966.2006.10302.x}
}

@ARTICLE{Edwards06,
       author = {{Edwards}, R. T. and {Hobbs}, G. B. and {Manchester}, R. N.},
        title = "{TEMPO2, a new pulsar timing package -- II. The timing model and precision estimates}",
      journal = {MNRAS},
         year = 2006,
        volume = {372},
        number = {4},
        pages = {1549--1574},
          doi = {10.1111/j.1365-2966.2006.10870.x},
}

@ARTICLE{Liddle07,
       author = {{Liddle}, Andrew R.},
        title = "{Information criteria for astrophysical model selection}",
      journal = {\mnras},
         year = 2007,
        month = may,
       volume = {377},
       number = {1},
        pages = {L74-L78},
          doi = {10.1111/j.1745-3933.2007.00306.x}
}

@ARTICLE{pdr14,
       author = {{Pennucci}, Timothy T. and {Demorest}, Paul B. and {Ransom}, Scott M.},
        title = "{Elementary Wideband Timing of Radio Pulsars}",
      journal = {\apj},
         year = 2014,
        month = aug,
       volume = {790},
       number = {2},
          eid = {93},
        pages = {93},
          doi = {10.1088/0004-637X/790/2/93}
}

@ARTICLE{ZSD15,
       author = {{Zhu}, WW and {Stairs}, IH and {Demorest}, PB and {et al.}},
        title = "{TESTING THEORIES OF GRAVITATION USING 21-YEAR TIMING OF PULSAR BINARY J1713+0747}",
      journal = {\apj},
         year = 2015,
        volume = {809(1)},
        pages = {41},
          doi = {10.1088/0004-637X/809/1/41}
}

@ARTICLE{ABB16,
       author = {{Arzoumanian}, Z and {Brazier}, A and {Burke-Spolaor}, S and {et al.}},
        title = "{THE NANOGRAV NINE-YEAR DATA SET: LIMITS ON THE ISOTROPIC STOCHASTIC GRAVITATIONAL WAVE BACKGROUND}",
      journal = {\apj},
         year = 2016,
        volume = {821(1)},
        pages = {13},
          doi = {10.3847/0004-637X/821/1/13}
}

@ARTICLE{css16,
       author = {{Cordes}, J.~M. and {Shannon}, R.~M. and {Stinebring}, D.~R.},
        title = "{Frequency-dependent Dispersion Measures and Implications for Pulsar Timing}",
      journal = {\apj},
         year = 2016,
        month = jan,
       volume = {817},
       number = {1},
          eid = {16},
        pages = {16},
          doi = {10.3847/0004-637X/817/1/16}
}

@ARTICLE{Lentati17,
       author = {{Lentati}, L. and {Kerr}, M. and {Dai}, S. and {Hobson}, M.~P. and {Shannon}, R.~M. and {Hobbs}, G. and {Bailes}, M. and {Bhat}, N.~D. Ramesh and {Burke-Spolaor}, S. and {Coles}, W. and {Dempsey}, J. and {Lasky}, P.~D. and {Levin}, Y. and {Manchester}, R.~N. and {Os{\l}owski}, S. and {Ravi}, V. and {Reardon}, D.~J. and {Rosado}, P.~A. and {Spiewak}, R. and {van Straten}, W. and {Toomey}, L. and {Wang}, J. and {Wen}, L. and {You}, X. and {Zhu}, X.},
        title = "{Wide-band profile domain pulsar timing analysis}",
      journal = {\mnras},
         year = 2017,
        month = apr,
       volume = {466},
       number = {3},
        pages = {3706-3727},
          doi = {10.1093/mnras/stw3359}
}

@ARTICLE{sc17,
       author = {{Shannon}, R.~M. and {Cordes}, J.~M.},
        title = "{Modelling and mitigating refractive propagation effects in precision pulsar timing observations}",
      journal = {\mnras},
         year = 2017,
        month = jan,
       volume = {464},
       number = {2},
        pages = {2075-2089},
          doi = {10.1093/mnras/stw2449}
}

@ARTICLE{Gupta17,
       author = {{Gupta}, Y. and {Ajithkumar}, B. and {Kale}, H.~S. and {Nayak}, S. and {Sabhapathy}, S. and {Sureshkumar}, S. and {Swami}, R.~V. and {Chengalur}, J.~N. and {Ghosh}, S.~K. and {Ishwara-Chandra}, C.~H. and {Joshi}, B.~C. and {Kanekar}, N. and {Lal}, D.~V. and {Roy}, S.},
        title = "{The upgraded GMRT: opening new windows on the radio Universe}",
      journal = {Current Science},
         year = 2017,
        month = aug,
       volume = {113},
       number = {4},
        pages = {707-714},
          doi = {10.18520/cs/v113/i04/707-714},
}

@ARTICLE{Reddy17,
       author = {{Reddy}, Suda Harshavardhan and {Kudale}, Sanjay and {Gokhale}, Upendra and {Halagalli}, Irappa and {Raskar}, Nilesh and {de}, Kishalay and {Gnanaraj}, Shelton and {Ajith Kumar}, B. and {Gupta}, Yashwant},
        title = "{A Wideband Digital Back-End for the Upgraded GMRT}",
      journal = {Journal of Astronomical Instrumentation},
         year = 2017,
        month = mar,
       volume = {6},
       number = {1},
          eid = {1641011-336},
        pages = {1641011-336},
          doi = {10.1142/S2251171716410117},
}

@ARTICLE{Joshi18,
       author = {{Joshi}, Bhal Chandra and {Arumugasamy}, Prakash and {Bagchi}, Manjari and {Bandyopadhyay}, Debades and {Basu}, Avishek and {Dhanda Batra}, Neelam and {Bethapudi}, Suryarao and {Choudhary}, Arpita and {De}, Kishalay and {Dey}, L. and {Gopakumar}, A. and {Gupta}, Y. and {Krishnakumar}, M.~A. and {Maan}, Yogesh and {Manoharan}, P.~K. and {Naidu}, Arun and {Nandi}, Rana and {Pathak}, Dhruv and {Surnis}, Mayuresh and {Susobhanan}, Abhimanyu},
        title = "{Precision pulsar timing with the ORT and the GMRT and its applications in pulsar astrophysics}",
      journal = {Journal of Astrophysics and Astronomy},
         year = 2018,
        month = aug,
       volume = {39},
       number = {4},
          eid = {51},
        pages = {51},
          doi = {10.1007/s12036-018-9549-y},
 }

@ARTICLE{Pennucci19,
       author = {{Pennucci}, Timothy T.},
        title = "{Frequency-dependent Template Profiles for High-precision Pulsar Timing}",
      journal = {\apj},
         year = 2019,
        month = jan,
       volume = {871},
       number = {1},
          eid = {34},
        pages = {34},
          doi = {10.3847/1538-4357/aaf6ef}
}

@article{Virtanen20,
  author  = {Virtanen, Pauli and Gommers, Ralf and Oliphant, Travis E. and
             Haberland, Matt and Reddy, Tyler and Cournapeau, David and
             Burovski, Evgeni and Peterson, Pearu and Weckesser, Warren and
             Bright, Jonathan and {et al.}},
  title   = {SciPy 1.0: Fundamental Algorithms for Scientific Computing in Python},
  journal = {Nature Methods},
  volume  = {17},
  number  = {3},
  pages   = {261--272},
  year    = {2020},
  doi     = {10.1038/s41592-019-0686-2}
}

@ARTICLE{SHML21,
       author = {{Shapiro-Albert}, BJ and {Hazboun}, JS and {McLaughlin}, MA and {Lam}, MT},
        title = "{A Study in Frequency-dependent Effects on Precision Pulsar Timing Parameters with the Pulsar Signal Simulator}",
      journal = {\apj},
         year = 2021,
        volume = {909:219},
        pages = {18},
          doi = {10.3847/1538-4357/abdc29}
}

@ARTICLE{Alam21,
       author = {{Alam}, Md F. and {Arzoumanian}, Zaven and {Baker}, Paul T. and {Blumer}, Harsha and {Bohler}, Keith E. and {Brazier}, Adam and {Brook}, Paul R. and {Burke-Spolaor}, Sarah and {Caballero}, Keeisi and {Camuccio}, Richard S. and {Chamberlain}, Rachel L. and {Chatterjee}, Shami and {Cordes}, James M. and {Cornish}, Neil J. and {Crawford}, Fronefield and {Cromartie}, H. Thankful and {Decesar}, Megan E. and {Demorest}, Paul B. and {Dolch}, Timothy and {Ellis}, Justin A. and {Ferdman}, Robert D. and {Ferrara}, Elizabeth C. and {Fiore}, William and {Fonseca}, Emmanuel and {Garcia}, Yhamil and {Garver-Daniels}, Nathan and {Gentile}, Peter A. and {Good}, Deborah C. and {Gusdorff}, Jordan A. and {Halmrast}, Daniel and {Hazboun}, Jeffrey S. and {Islo}, Kristina and {Jennings}, Ross J. and {Jessup}, Cody and {Jones}, Megan L. and {Kaiser}, Andrew R. and {Kaplan}, David L. and {Kelley}, Luke Zoltan and {Key}, Joey Shapiro and {Lam}, Michael T. and {Lazio}, T. Joseph W. and {Lorimer}, Duncan R. and {Luo}, Jing and {Lynch}, Ryan S. and {Madison}, Dustin R. and {Maraccini}, Kaleb and {McLaughlin}, Maura A. and {Mingarelli}, Chiara M.~F. and {Ng}, Cherry and {Nguyen}, Benjamin M.~X. and {Nice}, David J. and {Pennucci}, Timothy T. and {Pol}, Nihan S. and {Ramette}, Joshua and {Ransom}, Scott M. and {Ray}, Paul S. and {Shapiro-Albert}, Brent J. and {Siemens}, Xavier and {Simon}, Joseph and {Spiewak}, Ren{\'e}e and {Stairs}, Ingrid H. and {Stinebring}, Daniel R. and {Stovall}, Kevin and {Swiggum}, Joseph K. and {Taylor}, Stephen R. and {Tripepi}, Michael and {Vallisneri}, Michele and {Vigeland}, Sarah J. and {Witt}, Caitlin A. and {Zhu}, Weiwei and {Nanograv Collaboration}},
        title = "{The NANOGrav 12.5 yr Data Set: Wideband Timing of 47 Millisecond Pulsars}",
      journal = {\apjs},
         year = 2021,
        month = jan,
       volume = {252},
       number = {1},
          eid = {5},
        pages = {5},
          doi = {10.3847/1538-4365/abc6a1}
}

@ARTICLE{Joshi22,
       author = {{Joshi}, Bhal Chandra and {Gopakumar}, Achamveedu and {Pandian}, Arul and {Prabu}, Thiagaraj and {Dey}, Lankeswar and {Bagchi}, Manjari and {Desai}, Shantanu and {Tarafdar}, Pratik and {Rana}, Prerna and {Maan}, Yogesh and {Batra}, Neelam Dhanda and {Girgaonkar}, Raghav and {Agarwal}, Nikita and {Arumugam}, Paramasivan and {Basu}, Avishek and {Bathula}, Adarsh and {Dandapat}, Subhajit and {Gupta}, Yashwant and {Hisano}, Shinnosuke and {Kato}, Ryo and {Kharbanda}, Divyansh and {Kikunaga}, Tomonosuke and {Kolhe}, Neel and {Krishnakumar}, M.~A. and {Manoharan}, P.~K. and {Marmat}, Piyush and {Naidu}, Arun and {Banik}, Sarmistha and {Nobleson}, K. and {Paladi}, Avinash Kumar and {Pathak}, Dhruv and {Singha}, Jaikhomba and {Srivastava}, Aman and {Surnis}, Mayuresh and {Susarla}, Sai Chaitanya and {Susobhanan}, Abhimanyu and {Takahashi}, Keitaro},
        title = "{Nanohertz gravitational wave astronomy during SKA era: An InPTA perspective}",
      journal = {Journal of Astrophysics and Astronomy},
         year = 2022,
        month = dec,
       volume = {43},
       number = {2},
          eid = {98},
        pages = {98},
          doi = {10.1007/s12036-022-09869-w},
}

@ARTICLE{Nobleson22,
       author = {{Nobleson}, K. and {Agarwal}, Nikita and {Girgaonkar}, Raghav and {Pandian}, Arul and {Joshi}, Bhal Chandra and {Krishnakumar}, M.~A. and {Susobhanan}, Abhimanyu and {Desai}, Shantanu and {Prabu}, T. and {Bathula}, Adarsh and {Pennucci}, Timothy T. and {Banik}, Sarmistha and {Bagchi}, Manjari and {Dhanda Batra}, Neelam and {Choudhary}, Arpita and {Dandapat}, Subhajit and {Dey}, Lankeswar and {Gupta}, Yashwant and {Hisano}, Shinnosuke and {Kato}, Ryo and {Kharbanda}, Divyansh and {Kikunaga}, Tomonosuke and {Kolhe}, Neel and {Maan}, Yogesh and {Marmat}, Piyush and {Arumugam}, P. and {Manoharan}, P.~K. and {Pathak}, Dhruv and {Singha}, Jaikhomba and {Surnis}, Mayuresh P. and {Susarla}, Sai Chaitanya and {Takahashi}, Keitaro},
        title = "{Low-frequency wideband timing of InPTA pulsars observed with the uGMRT}",
      journal = {\mnras},
         year = 2022,
        month = may,
       volume = {512},
       number = {1},
        pages = {1234-1243},
          doi = {10.1093/mnras/stac532}
}

@ARTICLE{Tarafdar22,
       author = {{Tarafdar}, Pratik and {Nobleson}, K. and {Rana}, Prerna and {Singha}, Jaikhomba and {Krishnakumar}, M.~A. and {Joshi}, Bhal Chandra and {Paladi}, Avinash Kumar and {Kolhe}, Neel and {Batra}, Neelam Dhanda and {Agarwal}, Nikita and {Bathula}, Adarsh and {Dandapat}, Subhajit and {Desai}, Shantanu and {Dey}, Lankeswar and {Hisano}, Shinnosuke and {Ingale}, Prathamesh and {Kato}, Ryo and {Kharbanda}, Divyansh and {Kikunaga}, Tomonosuke and {Marmat}, Piyush and {Pandian}, B. Arul and {Prabu}, T. and {Srivastava}, Aman and {Surnis}, Mayuresh and {Susarla}, Sai Chaitanya and {Susobhanan}, Abhimanyu and {Takahashi}, Keitaro and {Arumugam}, P. and {Bagchi}, Manjari and {Banik}, Sarmistha and {De}, Kishalay and {Girgaonkar}, Raghav and {Gopakumar}, A. and {Gupta}, Yashwant and {Maan}, Yogesh and {Manoharan}, P.~K. and {Naidu}, Arun and {Pathak}, Dhruv},
        title = "{The Indian Pulsar Timing Array: First data release}",
      journal = {\pasa},
         year = 2022,
        month = oct,
       volume = {39},
          eid = {e053},
        pages = {e053},
          doi = {10.1017/pasa.2022.46}
}

@ARTICLE{Agazie23,
       author = {{Agazie}, Gabriella and {Alam}, Md Faisal and {Anumarlapudi}, Akash and {Archibald}, Anne M. and {Arzoumanian}, Zaven and {Baker}, Paul T. and {Blecha}, Laura and {Bonidie}, Victoria and {Brazier}, Adam and {Brook}, Paul R. and {Burke-Spolaor}, Sarah and {B{\'e}csy}, Bence and {Chapman}, Christopher and {Charisi}, Maria and {Chatterjee}, Shami and {Cohen}, Tyler and {Cordes}, James M. and {Cornish}, Neil J. and {Crawford}, Fronefield and {Cromartie}, H. Thankful and {Crowter}, Kathryn and {Decesar}, Megan E. and {Demorest}, Paul B. and {Dolch}, Timothy and {Drachler}, Brendan and {Ferrara}, Elizabeth C. and {Fiore}, William and {Fonseca}, Emmanuel and {Freedman}, Gabriel E. and {Garver-Daniels}, Nate and {Gentile}, Peter A. and {Glaser}, Joseph and {Good}, Deborah C. and {G{\"u}ltekin}, Kayhan and {Hazboun}, Jeffrey S. and {Jennings}, Ross J. and {Jessup}, Cody and {Johnson}, Aaron D. and {Jones}, Megan L. and {Kaiser}, Andrew R. and {Kaplan}, David L. and {Kelley}, Luke Zoltan and {Kerr}, Matthew and {Key}, Joey S. and {Kuske}, Anastasia and {Laal}, Nima and {Lam}, Michael T. and {Lamb}, William G. and {Lazio}, T. Joseph W. and {Lewandowska}, Natalia and {Lin}, Ye and {Liu}, Tingting and {Lorimer}, Duncan R. and {Luo}, Jing and {Lynch}, Ryan S. and {Ma}, Chung-Pei and {Madison}, Dustin R. and {Maraccini}, Kaleb and {McEwen}, Alexander and {McKee}, James W. and {McLaughlin}, Maura A. and {McMann}, Natasha and {Meyers}, Bradley W. and {Mingarelli}, Chiara M.~F. and {Mitridate}, Andrea and {Ng}, Cherry and {Nice}, David J. and {Ocker}, Stella Koch and {Olum}, Ken D. and {Panciu}, Elisa and {Pennucci}, Timothy T. and {Perera}, Benetge B.~P. and {Pol}, Nihan S. and {Radovan}, Henri A. and {Ransom}, Scott M. and {Ray}, Paul S. and {Romano}, Joseph D. and {Salo}, Laura and {Sardesai}, Shashwat C. and {Schmiedekamp}, Carl and {Schmiedekamp}, Ann and {Schmitz}, Kai and {Shapiro-Albert}, Brent J. and {Siemens}, Xavier and {Simon}, Joseph and {Siwek}, Magdalena S. and {Stairs}, Ingrid H. and {Stinebring}, Daniel R. and {Stovall}, Kevin and {Susobhanan}, Abhimanyu and {Swiggum}, Joseph K. and {Taylor}, Stephen R. and {Turner}, Jacob E. and {Unal}, Caner and {Vallisneri}, Michele and {Vigeland}, Sarah J. and {Wahl}, Haley M. and {Wang}, Qiaohong and {Witt}, Caitlin A. and {Young}, Olivia and {Nanograv Collaboration}},
        title = "{The NANOGrav 15 yr Data Set: Observations and Timing of 68 Millisecond Pulsars}",
      journal = {\apjl},
         year = 2023,
        month = jul,
       volume = {951},
       number = {1},
          eid = {L9},
        pages = {L9},
          doi = {10.3847/2041-8213/acda9a}
}

@ARTICLE{Curylo23,
       author = {{Cury{\l}o}, Ma{\l}gorzata and {Pennucci}, Timothy T. and {Bailes}, Matthew and {Bhat}, N.~D. Ramesh and {Cameron}, Andrew D. and {Dai}, Shi and {Hobbs}, George and {Kapur}, Agastya and {Manchester}, Richard N. and {Mandow}, Rami and {Miles}, Matthew T. and {Russell}, Christopher J. and {Reardon}, Daniel J. and {Shannon}, Ryan M. and {Spiewak}, Ren{\'e}e and {van Straten}, Willem and {Zhu}, Xing-Jiang and {Zic}, Andrew},
        title = "{Wide-band Timing of the Parkes Pulsar Timing Array UWL Data}",
      journal = {\apj},
         year = 2023,
        month = feb,
       volume = {944},
       number = {2},
          eid = {128},
        pages = {128},
          doi = {10.3847/1538-4357/aca535}
}

@ARTICLE{Miles23,
       author = {{Miles}, M.~T. and {Shannon}, R.~M. and {Bailes}, M. and {Reardon}, D.~J. and {Keith}, M.~J. and {Cameron}, A.~D. and {Parthasarathy}, A. and {Shamohammadi}, M. and {Spiewak}, R. and {van Straten}, W. and {Buchner}, S. and {Camilo}, F. and {Geyer}, M. and {Karastergiou}, A. and {Kramer}, M. and {Serylak}, M. and {Theureau}, G. and {Venkatraman Krishnan}, V.},
        title = "{The MeerKAT Pulsar Timing Array: first data release}",
      journal = {\mnras},
         year = 2023,
        month = mar,
       volume = {519},
       number = {3},
        pages = {3976-3991},
          doi = {10.1093/mnras/stac3644}
}

@article{Zic23,
    author        = {Zic, Andrew and Reardon, Daniel J. and
                     Kapur, Agastya and others},
    title         = {The Parkes Pulsar Timing Array third data release},
    journal       = {Publications of the Astronomical Society of Australia},
    volume        = {40},
    pages         = {e049},
    year          = {2023},
    doi           = {10.1017/pasa.2023.36},
    archivePrefix = {arXiv},
    eprint        = {2306.16230},
    primaryClass  = {astro-ph.HE}
}

@ARTICLE{Paladi24,
       author = {{Paladi}, Avinash Kumar and {Dwivedi}, Churchil and {Rana}, Prerna and {Nobleson}, K. and {Susobhanan}, Abhimanyu and {Joshi}, Bhal Chandra and {Tarafdar}, Pratik and {Deb}, Debabrata and {Arumugam}, Swetha and {Gopakumar}, A. and {Krishnakumar}, M.~A. and {Batra}, Neelam Dhanda and {Debnath}, Jyotijwal and {Kareem}, Fazal and {Arumugam}, Paramasivan and {Bagchi}, Manjari and {Bathula}, Adarsh and {Dandapat}, Subhajit and {Desai}, Shantanu and {Gupta}, Yashwant and {Hisano}, Shinnosuke and {Kharbanda}, Divyansh and {Kikunaga}, Tomonosuke and {Kolhe}, Neel and {Maan}, Yogesh and {Manoharan}, P.~K. and {Singha}, Jaikhomba and {Srivastava}, Aman and {Surnis}, Mayuresh and {Takahashi}, Keitaro},
        title = "{Multiband extension of the wideband timing technique}",
      journal = {\mnras},
         year = 2024,
        month = jan,
       volume = {527},
       number = {1},
        pages = {213-231},
          doi = {10.1093/mnras/stad3122}
}

@ARTICLE{Susobhanan24,
       author = {{Susobhanan}, A. and {Kaplan}, D. L. and {Archibald}, A. M. and
                 {Luo}, J. and {Ray}, P. S. and {Pennucci}, T. T. and
                 {Ransom}, S. M. and {Agazie}, G. and {Fiore}, W. and
                 {Larsen}, B. and {O'Neill}, P. and {van Haasteren}, R. and
                 {Anumarlapudi}, A. and {Bachetti}, M. and {Bhakta}, D. and
                 {Champagne}, C. A. and {Cromartie}, H. T. and
                 {Demorest}, P. B. and {Jennings}, R. J. and {Kerr}, M. and
                 {Levina}, S. and {McEwen}, A. and {Shapiro-Albert}, B. J. and
                 {Swiggum}, J. K.},
        title = "{PINT: Maximum-likelihood Estimation of Pulsar Timing Noise Parameters}",
      journal = {ApJ},
         year = 2024,
        volume = 971,
        number = 2,
       eid = {150},
          doi = {10.3847/1538-4357/ad59f7},
}

@ARTICLE{Rana25,
       author = {{Rana}, Prerna and {Tarafdar}, Pratik and {Nobleson}, K. and {Dwivedi}, Churchil and {Chandra Joshi}, Bhal and {Deb}, Debabrata and {Mondal}, Sushovan and {Krishnakumar}, M.~A. and {Shukla}, Adya and {Singha}, Jaikhomba and {Grover}, Himanshu and {Tahbildar}, Hemanga and {Susobhanan}, Abhimanyu and {Surnis}, Mayuresh and {Desai}, Shantanu and {Batra}, Neelam Dhanda and {Srivastava}, Aman and {Bharambe}, Vinay and {Jose}, Jibin and {Vyasraj}, Vaishnavi and {Jose Jacob}, Shebin and {Amarnath} and {Singh}, Manpreet and {Zuraiq}, Zenia and {Sengupta}, Sarbartha and {Ogi}, Toki and {Kumar}, Dhruv and {Jagadeesh}, S. and {Kareem}, Fazal and {Maity}, Deep and {Rai}, Kaustubh and {Vara}, Kunjal and {Chowdhury}, Shaswata and {Kato}, Ryo and {Arumugam}, Swetha and {Mamidipaka}, Pragna and {Arul Pandian}, B. and {Shaji}, Kavya and {Thiagaraj}, Prabu and {Arumugam}, Paramasivan and {Bagchi}, Manjari and {Chakraborty}, Manoneeta and {Gopakumar}, Achamveedu and {Gupta}, Yashwant and {Maan}, Yogesh and {Kumar Paladi}, Avinash and {Takahashi}, Keitaro},
        title = "{The Indian Pulsar Timing Array data release 2: I. Dataset and timing analysis}",
      journal = {\pasa},
         year = 2025,
        month = jul,
       volume = {42},
          eid = {e108},
        pages = {e108},
          doi = {10.1017/pasa.2025.10066}
}

@ARTICLE{Miles25,
       author = {{Miles}, Matthew T. and {Shannon}, Ryan M. and {Reardon}, Daniel J. and {Bailes}, Matthew and {Champion}, David J. and {Geyer}, Marisa and {Gitika}, Pratyasha and {Grunthal}, Kathrin and {Keith}, Michael J. and {Kramer}, Michael and {Kulkarni}, Atharva D. and {Nathan}, Rowina S. and {Parthasarathy}, Aditya and {Porayko}, Nataliya K. and {Singha}, Jaikhomba and {Theureau}, Gilles and {Abbate}, Federico and {Buchner}, Sarah and {Cameron}, Andrew D. and {Camilo}, Fernando and {Moreschi}, Beatrice E. and {Shaifullah}, Golam and {Shamohammadi}, Mohsen and {Krishnan}, Vivek Venkatraman},
        title = "{The MeerKAT Pulsar Timing Array: the 4.5-yr data release and the noise and stochastic signals of the millisecond pulsar population}",
      journal = {\mnras},
         year = 2025,
        month = jan,
       volume = {536},
       number = {2},
        pages = {1467-1488},
          doi = {10.1093/mnras/stae2572}
}

@ARTICLE{Susobhanan26,
       author = {{Susobhanan}, Abhimanyu and {Paladi}, Avinash Kumar and {Desmecht}, R{\'e}ka and {Amarnath} and {Bagchi}, Manjari and {Chakraborty}, Manoneeta and {Chowdhury}, Shaswata and {Das}, Suruj Jyoti and {Deb}, Debabrata and {Desai}, Shantanu and {Dwivedi}, Churchil and {Grover}, Himanshu and {Jose}, Jibin and {Joshi}, Bhal Chandra and {Kala}, Shubham and {Kareem}, Fazal and {Meena}, Kuldeep and {Mondal}, Sushovan and {Nobleson}, K. and {Pandian}, B. Arul and {Rai}, Kaustubh and {Shukla}, Adya and {Singh}, Manpreet and {Srivastava}, Aman and {Surnis}, Mayuresh and {Tahbildar}, Hemanga and {Takahashi}, Keitaro and {Tarafdar}, Pratik and {Vara}, Kunjal and {Vyasraj}, Vaishnavi and {Zuraiq}, Zenia},
        title = "{Revisiting wideband pulsar timing measurements}",
      journal = {\mnras},
         year = 2026,
        month = apr,
       volume = {547},
       number = {4},
          eid = {stag444},
        pages = {stag444},
          doi = {10.1093/mnras/stag444}
}

@ARTICLE{sofia26,
      title={The NANOGrav 15 yr Dataset: Modified Functional Forms for Profile-Evolution Time Delays in Pulsar Timing Models}, 
      author={{Sosa Fiscella}, Sofia V.  and Michael T. Lam and Olivia West and Ruggero Valdata and Gabriella Agazie and Akash Anumarlapudi and Anne M. Archibald and Zaven Arzoumanian and Paul T. Baker and Paul R. Brook and H. Thankful Cromartie and Kathryn Crowter and Megan E. DeCesar and Paul B. Demorest and Timothy Dolch and Elizabeth C. Ferrara and William Fiore and Emmanuel Fonseca and Gabriel E. Freedman and Nate Garver-Daniels and Peter A. Gentile and Joseph Glaser and Deborah C. Good and Jeffrey S. Hazboun and Ross J. Jennings and Megan L. Jones and David L. Kaplan and Matthew Kerr and Duncan R. Lorimer and Jing Luo and Ryan S. Lynch and Alexander McEwen and Maura A. McLaughlin and Natasha McMann and Bradley W. Meyers and Cherry Ng and David J. Nice and Timothy T. Pennucci and Benetge B. P. Perera and Nihan S. Pol and Henri A. Radovan and Scott M. Ransom and Paul S. Ray and Ann Schmiedekamp and Carl Schmiedekamp and Brent J. Shapiro-Albert and Taylor Starkman and Ingrid H. Stairs and Kevin Stovall and Abhimanyu Susobhanan and Joseph K. Swiggum and Haley M. Wahl},
      year={2026},
      eprint={2610.10431},
      archivePrefix={arXiv},
      primaryClass={astro-ph.HE},
      url={https://arxiv.org/abs/2610.10431}, 
}

@article{DD26,
	author = {{Deb, D.} and {Tarafdar, P.} and {Joshi, B. C.} and {Bagchi, M.} and {Chowdhury, S.} and {Dwivedi, C.} and {Grover, H.} and {Amarnath}},
	title = {Multi-statistic validation of frequency-resolved templates for mitigating FD parameter requirements in narrowband pulsar timing},
	year = {2026},
}

\end{document}